\documentclass[aps,prl,twocolumn,superscriptaddress,groupedaddress]{revtex4}  
\usepackage{graphicx}  
\usepackage{dcolumn}   
\usepackage{bm}        
\usepackage{amssymb}   
\usepackage{amsmath}
\usepackage{color}
\usepackage{placeins}
\usepackage{hyperref}

\begin{document}


\title{Sequential dispersive measurement of a superconducting qubit}

\author{T. Peronnin}
\affiliation{Universit\'e Lyon, ENS de Lyon, Universit\'e Claude Bernard Lyon 1, CNRS, Laboratoire de Physique, F-69342
Lyon, France}

\author{D. Markovi\'c}
\affiliation{Unit\'e Mixte de Physique, CNRS, Thales, Univ. Paris-Sud, Universit\'e Paris-Saclay, 91767 Palaiseau,
France
}

\author{Q. Ficheux}

\affiliation{Universit\'e Lyon, ENS de Lyon, Universit\'e Claude Bernard Lyon 1, CNRS, Laboratoire de Physique, F-69342
Lyon, France}

\author{B. Huard}
\affiliation{Universit\'e Lyon, ENS de Lyon, Universit\'e Claude Bernard Lyon 1, CNRS, Laboratoire de Physique, F-69342
Lyon, France}

\date{\today}

\begin{abstract}
We present a superconducting device that realizes the sequential measurement of a transmon qubit. The device disables common limitations of dispersive readout such as Purcell effect or transients in the cavity mode by turning on and off the coupling to the measurement channel on demand. The qubit measurement begins by loading a readout resonator that is coupled to the qubit. After an optimal interaction time with negligible loss, a microwave pump releases the content of the readout mode by upconversion into a measurement line in a characteristic time as low as 10~ns, which is 400 times shorter than the lifetime of the readout resonator. A direct measurement of the released field quadratures demonstrates a readout fidelity of $97.5~\%$ in a total measurement time of $220~\mathrm{ns}$. The Wigner tomography of the readout mode allows us to characterize the non-Gaussian nature of the readout mode and its dynamics.
\end{abstract}

\maketitle


One of the main differences between quantum and classical physics lies in the fact that a measurement inherently disturbs a quantum system. When the measurement does not destroy the system (Quantum Non Demolition measurement or QND), it leads to a backaction that updates its wavefunction. A basic measurement model introduces a probe, which is an ancillary quantum system. The probe is prepared in a given state before interacting with the system under scrutiny for a time $t_\mathrm{int}$, which is able to generate an entangled state between probe and system. After this pre-measurement step, the probe is sent to a detector. A random outcome is selected, which leads to the collapse of the system into the state corresponding to that outcome~\cite{Zurek2003}. 
\begin{figure}[!h]
\includegraphics[scale=0.95]{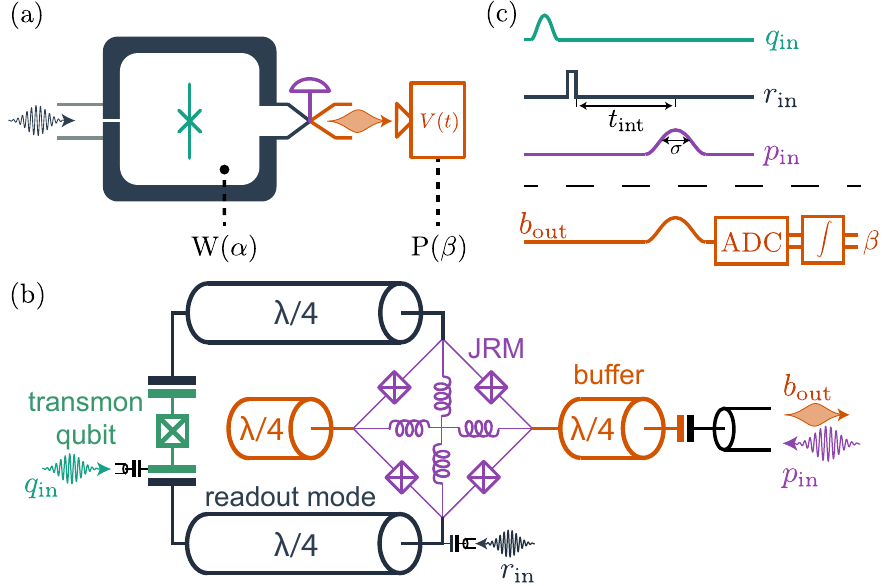}
\caption{(a) The readout resonator (dark gray) is first prepared in a coherent state to probe a transmon qubit. Owing to the dispersive interaction between them, the resonator and the qubit evolve unitarily during a time $t_\mathrm{int}$. The readout mode is then upconverted and released into an output line (via purple valve). The complex amplitude $\beta$ of the released mode is encoded in the recorded output voltage $V(t)$. Both the Wigner function $W(\alpha)$ of the readout mode and the probability distribution $P(\beta)$ can be measured. (b) Scheme of the device. The readout mode is a $\lambda/2$ coplanar waveguide resonator (dark gray) that is capacitively coupled to a transmon qubit (green). The readout resonator is coupled to another $\lambda/2$ resonator (buffer in orange) by a Josephson Ring Modulator (JRM in purple). The readout resonator is driven by $r_\mathrm{in}$, the transmon (green) by $q_\mathrm{in}$, the buffer resonator (orange) outputs in $b_\mathrm{out}$, and the JRM is pumped at amplitude $p_\mathrm{in}$ by the buffer input. (c) Pulse sequence of the qubit measurement (top) and the released field measurement scheme (bottom).}
\label{schema}
\end{figure}
The inner mechanics of that measurement process can be illustrated in Circuit quantum-electrodynamics (circuit-QED). Owing to the dispersive interaction, a driven stationary microwave mode can act as a probe of the state of a coupled superconducting qubit. The output of the stationary mode is recorded and leads to a continuous measurement record. The record can then be integrated in time to implement an effective single projective measurement~\cite{Blais2007}, or taken into account as a function of time to determine the quantum trajectory followed by the qubit~\cite{Gambetta2008,Murch2013a,Tan2014,Weber2016,ficheux2017dynamics} or even to realize measurement based feedback~\cite{Vijay2012c,Riste2012,Campagne-Ibarcq2013,Riste2013a,DeLange2014}. However, the three steps of the basic measurement process are simultaneous, as the probe gets refilled and leaks out information during the interaction time.

Here, we present a circuit-QED experiment, where the measurement of a qubit in the $\{|g\rangle,|e\rangle\}$ basis is separated in the three sequential steps of the basic measurement model. The device enables the qubit readout based on tunable cavity couplers that were proposed in the past~\cite{Sete2013,Gard2019}. For that purpose, we have designed and realized a transmon qubit coupled to a high-Q microwave readout resonator whose state can be flushed on-demand into an output line (see Fig.~\ref{schema}a). The release characteristic time can be as short as $10~\mathrm{ns}$, thus considerably improving a former 3D version of this device~\cite{Flurin2014}. The readout resonator is initialized in a coherent state and evolves unitarily in interaction with the qubit for a time $t_\mathrm{int}$. Finally the quantum state of the readout resonator is released into the transmission line and measured, thus revealing information about the qubit state. Interestingly, the scheme alleviates the usual trade-off between QND measurement speed and fidelity by disabling the link between measurement time and qubit relaxation rate~\cite{Reed,Jeffrey2014,Walter2017} and strongly shortening transients in resonator population~\cite{Jeffrey2014,McClure2015,Bultink2016}, without resorting to the complexity of longitudinal coupling~\cite{Didier2015,Touzard2018,Ikonen2018}. We obtain performances that are close to state-of-the-art for qubit readout~\cite{Walter2017} with a fidelity of 97.5~\% in a total time of 220~ns, which shows that our device could be an excellent technical choice once its coupling rates are optimized.

The key element of our sequential measurement is a readout mode that couples to a qubit and which is able to store a microwave field and release it on-demand into an output line. In order to realize it, we have used an upgraded version of the Superconducting Quantum Node~\cite{Flurin2014} by coupling it to a transmon qubit. Instead of using 3D architecture as in our previous work, the device (see Fig.~\ref{schema}b) is made in coplanar waveguide geometry (CPW) with Niobium on Silicon and Al/AlOx/Al Josephson junctions. The Quantum Node is made of two CPW $\lambda/2$ resonators that are coupled via a Josephson Ring Modulator (JRM) in their center and is cooled down below 30~mK. The readout resonator has a frequency $\omega_r = 2\pi \times3.73 ~ \mathrm{GHz}$ and a lifetime $T_r = 4.0~\mathrm{\mu s}$. The shorter buffer resonator has a frequency $\omega_b = 2\pi \times 10.22 ~ \mathrm{GHz}$ and is intentionally much more lossy as it is connected to an output transmission line with a rate $\kappa_b = 2\pi \times 21 ~ \mathrm{MHz}$. 

Close to flux quantum in the Josephson ring, the coupling Hamiltonian is dominated by a three-wave mixing interaction term between the readout mode ($\hat{r}$), the buffer mode ($\hat{b}$) and a common mode of the device that is localized in both arms~\cite{supmat}. When the latter is off-resonantly driven by a pump of amplitude $p_{in}$ and frequency $\omega_p = \omega_b - \omega_r$, the effective coupling is described by the beam-splitter Hamiltonian $\hat{H}_{\mathrm{bs}} = \hbar (g \hat{b}^\dagger \hat{r} + g^* \hat{b} \hat{r}^\dagger)$, where the conversion rate $g$ is proportional to the pump amplitude $p_{in}$~\cite{Abdo2013a}. With the addition of a frequency conversion between readout and buffer modes, turning on the pump thus results in an effective tunable coupling rate of the readout mode to the output line~\cite{Flurin2014,Pfaff2017a,supmat}. In practice, we used a smoothed time dependence of the pump amplitude $p_\mathrm{in}(t) \propto \left(\cosh{(\sqrt{\pi/2} t/ \sigma)}\right)^{-1}$ in order to turn on or off the coupling (Fig.~\ref{schema}c). While the readout mode can be emptied in a timescale as short as $\sigma_\mathrm{min} = 10 ~ \mathrm{ns}$ \cite{supmat}, we obtained the highest measurement fidelity for a  time $\sigma = 28 ~ \mathrm{ns}$, and we consider this case only in the rest of the letter.
\begin{figure*}
\includegraphics[scale=1.0]{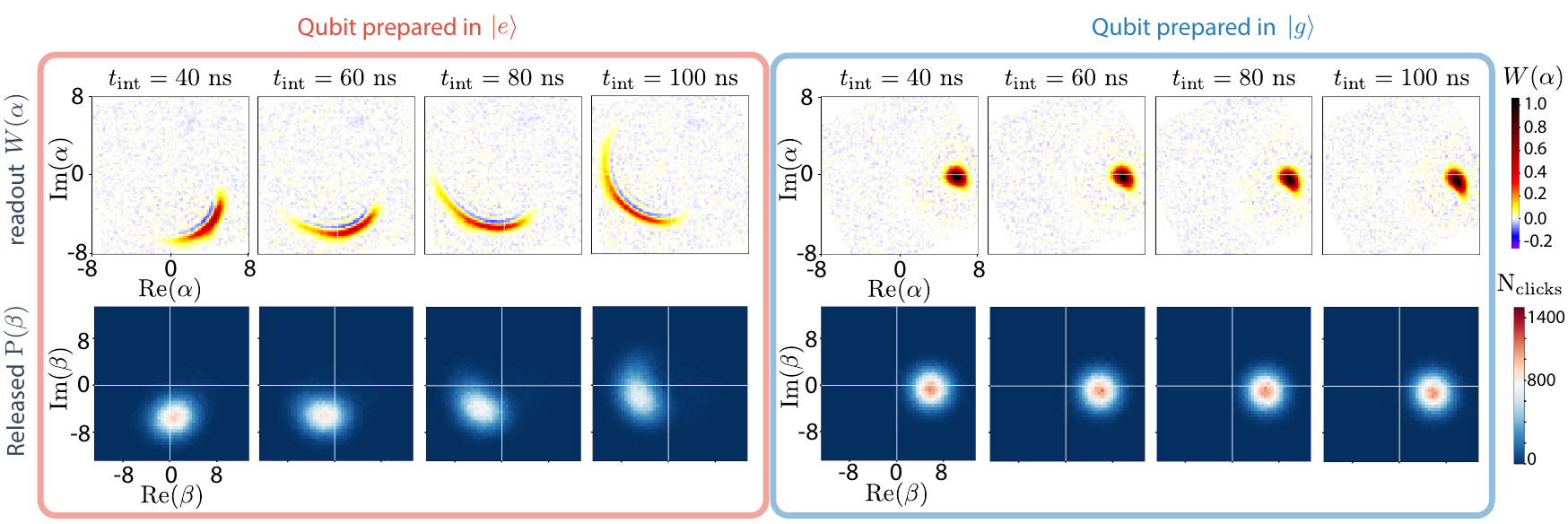}
\caption{Top row: measured Wigner function $W(\alpha)$ of the readout mode after various interaction times $t_\mathrm{int}$ for a qubit initialized in $|e\rangle$ or $|g\rangle$ and initial coherent state amplitude $\alpha_0 = 5.8$. A global rotation of the quadrature phase space was done numerically~\cite{supmat}. Bottom row: corresponding histograms of the measured complex amplitude $\beta$ of the output field (see text for definition) after it has been released into the transmission line following the interaction time $t_\mathrm{int}$ for $10^5$ realizations of the experiment.}
\label{Tomo}
\end{figure*}

The system to measure is a transmon qubit with frequency $\omega_q = 2\pi \times4.45 ~ \mathrm{GHz}$, lifetime $T_1 = 6.1~\mathrm{\mu s}$ and coherence time $T_2 = 9.2~\mathrm{\mu s}$. Crucially, the symmetry of the superconducting circuit is chosen to cancel out its coupling to the common mode while preserving the coupling to the $\lambda/2$ mode of the readout resonator. Therefore the large pump powers have no detrimental effect on the transmon during release. A previous design without this symmetry showed that the pump would otherwise ionize the transmon~\cite{supmat,Lescanne2018a}. The qubit and readout modes are dispersively coupled so that the effective first order interaction term reads $-\hbar\chi \hat{r}^\dagger \hat{r}|e\rangle \langle e|$ where $\chi = 2\pi \times 2.1 ~\mathrm{MHz}$. Once the qubit is prepared in a given state (Fig.~\ref{schema}c), the experiment starts by a displacement of the initially empty readout mode (via port $r_\mathrm{in}$) in $10~\mathrm{ns}$, which is short enough compared to $1/\chi$ to prepare the readout mode in the coherent state $|\alpha_0\rangle$ irrespective of the qubit state. The measurement probe is thus initialized in a deterministic state and can now interact with the qubit under a unitary evolution for a time $t_\mathrm{int}$. After this time, the content of the readout mode is released into the transmission line by upconversion to the buffer frequency using a pump pulse. It is finally amplified and measured to extract its complex amplitude $\beta$ as defined below.

Before discussing how to retrieve the information on the qubit state from the final measurement of the released field, we can benefit from the sequential aspect of the process to experimentally investigate the dynamics of the readout mode during its interaction with the qubit. The dispersive interaction leads to a constant readout amplitude $\langle\hat{r}(t_\mathrm{int})\rangle=\alpha_0$ in the rotating frame at $\omega_r$ when the qubit is in $|g\rangle$, while its phase increases linearly in time $\langle\hat{r}(t_\mathrm{int})\rangle=\alpha_0 e^{i\chi t_\mathrm{int}}$ when in $|e\rangle$. This average behavior can be seen as an apparent rotation of the measured Wigner functions $W(\alpha)$ in the quadrature phase space of the readout mode as interaction time increases (Fig.~\ref{Tomo}) only when the qubit is prepared in $|e\rangle$. The Wigner tomography is here realized using the qubit as a photon number parity detector as in previous works~\cite{Lutterbach1997,Bertet2002,Vlastakis2013,Bretheau2015,supmat}. 
 
In order to read out the state of the qubit as well as possible, it is desirable to maximize the distance between the states of the released microwave mode corresponding to $|g\rangle$ and $|e\rangle$. It was proposed in Ref.~\cite{Sete2013} that using the superconducting qubit non-linearity to generate squeezing could help reducing the overlap between the two states. From the measured Wigner functions (Fig.~\ref{Tomo}), it appears that the state of the readout mode is not a coherent state when the qubit is excited, as shown by a non Gaussian shape and by the development of negativities~\cite{Khezri2016}. The ability to realize a direct Wigner tomography of the readout mode, and the measurement of negativities, demonstrates that the measurement probe has not decohered entirely prior to the release into the transmission line. The quantum-classical boundary occurs at a later stage. The dynamics can be understood by expanding the interaction Hamiltonian at least to the next order~\cite{Kirchmair2013a}, which gives 
\begin{equation}
\hat{H}_{\textrm{int}} / \hbar = -\chi \hat{r}^\dagger \hat{r}|e\rangle \langle e|- K_g \hat{r}^{\dagger 2} \hat{r}^2 |g\rangle \langle g|- K_e \hat{r}^{\dagger 2} \hat{r}^2 |e\rangle \langle e|.
\label{Hint}
\end{equation}
The Kerr rates could be independently measured to be $K_g = 2\pi\times 8~\mathrm{kHz}$ and $K_e = 2\pi\times 37~\mathrm{kHz}$ ~\cite{supmat}. Interestingly, the self-Kerr rate of the readout mode is much larger when the qubit is excited than when it is in the ground state, which explains why the Wigner function shape is little distorted for $|g\rangle$ and strongly distorted for $|e\rangle$. A simple way to qualitatively understand the shape of the Wigner function is to realize that the Kerr term induces an angular velocity in the quadrature phase space that increases with field amplitude. A quantitative study reveals that yet higher order terms need to be taken into account at the large number of photons $|\alpha_0|^2\approx 34$ that we are using~\cite{supmat}. It is interesting to note that it would be possible to use a single junction for releasing the readout mode~\cite{Pfaff2017a}. However, the JRM we use allows us to tune the cross-Kerr terms by the magnetic flux in the ring~\cite{supmat}. In this work, we set the flux to cancel out the cross-Kerr effects between the buffer and the readout mode and between the readout mode and the common mode. Hence the strong pump does not shift the resonance frequencies.


\begin{figure}
\includegraphics[scale=1.0]{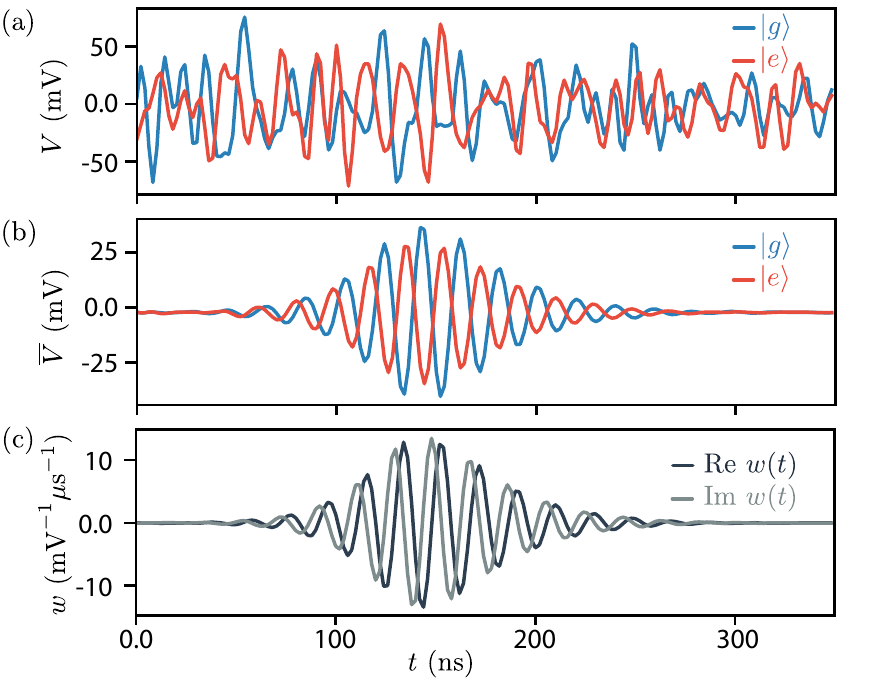}
\caption{ (a) Raw traces of the output voltage $V(t)$ for two realizations of the experiment with the qubit prepared either in $|g\rangle $ or in $|e\rangle$ and with an initial displacement $\alpha_0 = 5.8$ and interaction time $t_\mathrm{int} = 100~\textrm{ns}$. (b) Average traces over $10^5$ realizations.  (c) Real and imaginary part of the weight function $w(t)$ from which the complex amplitude of the released mode is obtained.
}
\label{records}
\end{figure}
By releasing the content of the readout mode into the output line by upconversion to $\omega_b$, we can access the information about the qubit that is encoded in the readout resonator. The emitted pulse is amplified by a Traveling Wave Parametric Amplifier~\cite{Macklin2015} followed by cryogenic and room temperature amplifiers~\cite{supmat}. It is finally down-converted to $50~\mathrm{MHz}$ and digitized as a voltage $V(t)$ (Fig.~\ref{schema}c). A typical trace $V(t)$ of a single realization  is shown in Fig.~\ref{records}a for a qubit prepared in $|g\rangle$ or $|e\rangle$. How to recover the information about the qubit state from the final measurement of the released 
field? 
Our solution consists in defining a complex amplitude $\beta$ for the released mode by demodulating the single measurement records $V(t)$ by a single complex weight function $w(t)$ to be determined. 
\begin{equation}
\beta = \int{V(t) w(t) dt}.
\label{a_tilde}
\end{equation}
We first average the traces of $10^5$ realizations of the experiment for $t_\mathrm{int}=100~\mathrm{ns}$ and $\alpha_0=5.8$ (Fig.~\ref{records}b). The averages $\overline{V}_{|e,g\rangle}(t)$ reveal the 50~MHz modulation we used and the temporal envelope of the releasing pump $p_\mathrm{in}(t)$. 
 We find (see~\cite{supmat} for details) that a way to faithfully map the intraresonator complex field amplitude $\alpha$ to the demodulated amplitude $\beta$, independently of the qubit state, consists in choosing a weight function $w(t)$ whose real part is $\mathrm{Re}[w(t)] = \left( \overline{V}_{|e\rangle}(t) - \overline{V}_{|g\rangle}(t)\right)/2 \lambda$ (Fig.~\ref{records} c). Its imaginary part $\mathrm{Im}[w(t)]$ is then constructed by shifting the phase of the carrier modulation by $\pi/2$. The prefactor $\lambda^{-1}$ is adjusted so that $\beta$ is dimensionless and equal on average to $\alpha_0$ in the case $t_\mathrm{int}=0$.
As can be seen in Fig.~\ref{Tomo}, with this choice, the probability densities $P_{g,e}(\beta)$ to measure a complex amplitude $\beta$ for a qubit prepared in $|g\rangle$ and in $|e\rangle$ are indeed smoothed versions of the intraresonator Wigner functions owing to the $11~\%$ efficiency of the detection setup~\cite{supmat}.

\begin{figure}
\includegraphics[width=\columnwidth]{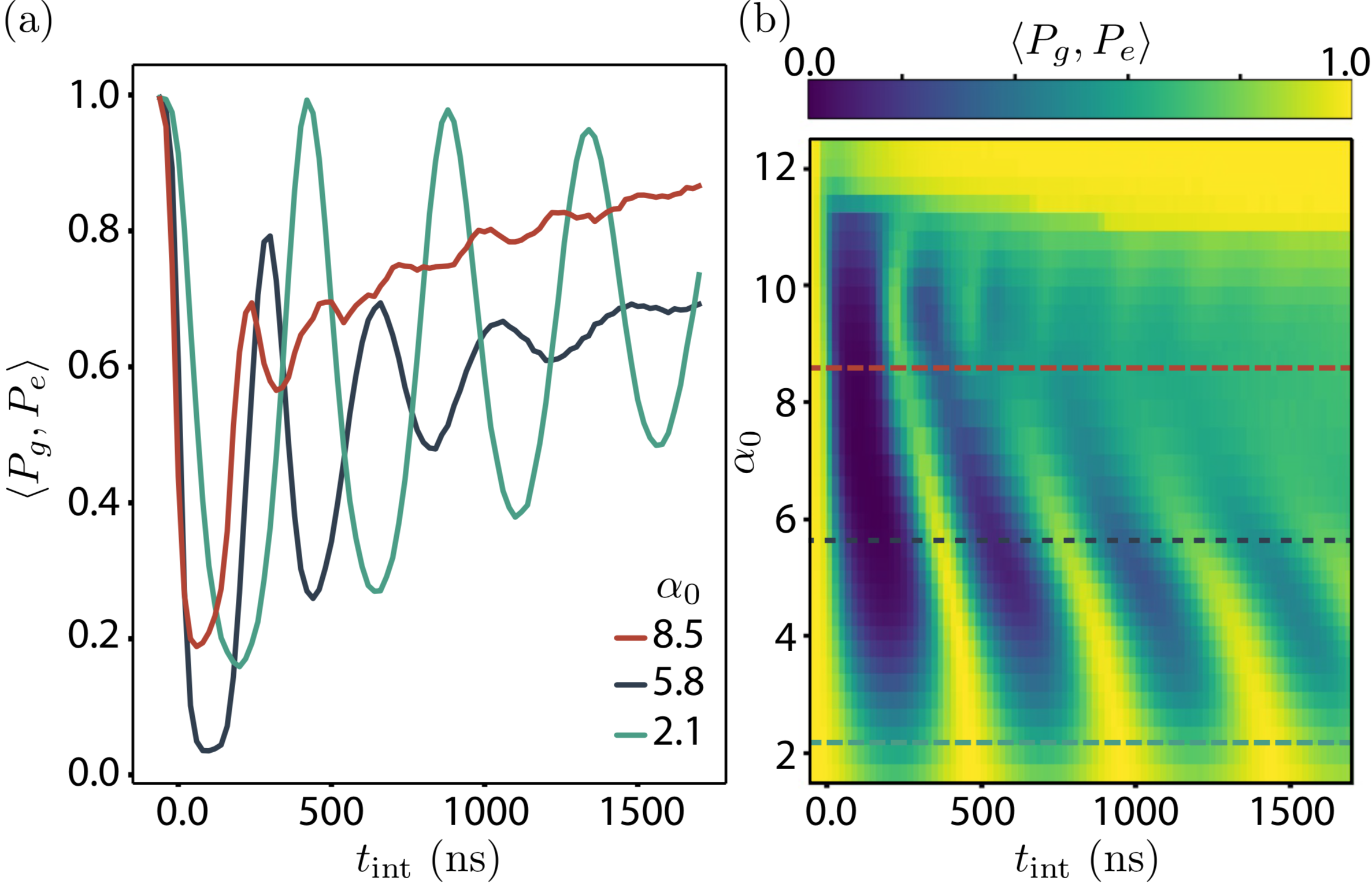}
\caption{ (a) Scalar product of the measured distributions $P_{g,e}(\beta)$ of released mode amplitudes when the qubit is prepared in $|g\rangle$ or $|e\rangle$ as a function of interaction time $t_\mathrm{int}$, and for various values of initial displacement $\alpha_0$. (b) Same scalar product as a function of $\alpha_0$ and $t_\mathrm{int}$. Colored dashed lines indicate cuts corresponding to Fig.~(a).
}
\label{overlap}
\end{figure}

The qubit measurement then comes down to determining whether a measured amplitude $\beta$ is more likely to occur if the qubit is in the ground or excited state. It is straightforward once we know the distributions of complex amplitudes $P_{g}(\beta)$ and $P_{e}(\beta)$. We stress that since the readout and released modes are not in a Gaussian state, there is information in both quadratures, which justifies the use of a phase preserving amplifier. The 2D scalar product between the histograms 
\begin{equation}
\left\langle P_{g},P_{e}\right\rangle=\frac{\int_\mathbf{C}\mathrm{d}\alpha P_g(\alpha)P_e(\alpha)}{\left(\int_\mathbf{C}\mathrm{d}\alpha P_g(\alpha)^2\right)^{1/2}\left(\int_\mathbf{C}\mathrm{d}\alpha P_e(\alpha)^2\right)^{1/2}}
\end{equation}
quantifies the distinguishability of the measurement outcomes. Fig.~\ref{overlap} represents the measured overlap of the two distributions as a function of interaction time $t_\mathrm{int}$ and for three values of the initial amplitude $\alpha_0$ of the readout mode. 

As mentioned above, the dispersive interaction leads to a global rotation in the quadrature phase space of the readout mode when the qubit is in $|e\rangle$. With dispersive interaction alone, the even and odd integer values of $t_\mathrm{int} \chi/\pi$ would thus correspond to maxima and minima of $\left\langle P_{g},P_{e}\right\rangle$, which are reached for full turns or half turns of the average complex amplitude of the readout mode. Fig.~\ref{overlap} illustrates which phenomena determine the amplitude $\alpha_0$ that maximizes the qubit measurement fidelity. If  $\alpha_0$ is too low, the separation between the distribution supports does not overcome the noise. It is what sets the minimum overlap of the green curve at $\alpha_0=2.1$, on top of the residual internal losses of the readout mode that dampen the oscillations. If  $\alpha_0$ is too large, the higher order terms in the interaction Hamiltonian lead to a uniform distribution in phase, and the time oscillations disappear (red curve at $\alpha_0=8.5$). The full dependence of the overlap $\left\langle P_{g},P_{e}\right\rangle$ as a function of $\alpha_0$ and $t_\mathrm{int}$ (Fig.~\ref{overlap}b) can be used to determine the optimal measurement conditions. The minimal overlap $\left\langle P_{g},P_{e}\right\rangle = 3.4~\%$ is reached at $\alpha_0 = 5.8$ and $t_{\mathrm{int}} = 100~\mathrm{ns}$ as in Fig.~\ref{Tomo} and \ref{records}. Interestingly, at large amplitudes ($7<\alpha_0<10$) the oscillations become more complicated as a double periodicity appears and is not understood yet (red curve). At very large values ($\alpha_0>10$) the qubit is ionized~\cite{Lescanne2018a}, which sets a stringent bound on the number of photons for quantum non-demolition readout. We note that this platform seems suited to study the dynamics of ionization of the qubit.

In order to characterize the measurement operation for applications, we associate values of $\beta$ where $P_{g}(\beta) > P_{e}(\beta) $ to the measurement outcome "g'' and "e'' otherwise. We can then define the measurement errors $\mathcal{E}_e$ (respect. $\mathcal{E}_g$) as obtaining the result "g" (respect. "e") after having prepared the qubit in $|e\rangle$ (respect. $|g\rangle$). We find $\mathcal{E}_e =~3.4\%$ and $\mathcal{E}_g =~1.6\%$. They can be partially explained by the qubit thermal population of $0.8 \%$, the finite qubit lifetime ($1.7\%$ in $\mathcal{E}_e$) and by the $99.5\%$ fidelity of the $\pi$ pulse from ground excited states~\cite{supmat}. The remaining arises from imperfect separation of the histograms. The average fidelity is $\mathcal{F} = 1 - \frac{\mathcal{E}_g + \mathcal{E}_e}{2} = 97.5~\%$ for a total qubit measurement time is 220~ns. Finally, one may wonder how the release process affects the qubit. We have characterized how QND the measurement is by determining the probability $\mathcal{F}_\mathrm{QND} = 95~\%$ that two successive measurements find the same outcome.


In conclusion, we have implemented the sequential measurement of a transmon where the probe initialization, interaction with the qubit and detection are all separated in time and space. Our readout scheme is insensitive to the Purcell effect since it is always coupled to a high Q resonator except during release when the buffer resonator acts as a Purcell filter anyway. By releasing the probing field on demand, we also avoid common limitations due to slow reset of the cavity mode in dispersive measurements. Further increasing both $\chi$ and $\kappa_b$ by an order of magnitude should straightforwardly lead to measurement times beyond state-of-the art. The use of a JRM as a switch between the readout mode and the transmission line opens interesting perspectives. Indeed, the JRM can also act as a built-in amplifier~\cite{Bergeal,Roch2012} (by applying a pump at the frequency $\omega_p = \omega_b + \omega_r$) to amplify the probing field as it is released into the transmission line. Besides, by increasing the participation ratio of the JRM in the readout mode, one could not only adjust but even cancel out the Kerr rates $K_g$ or $K_e$ of the readout mode by tuning the magnetic flux in the JRM. Furthermore it would be interesting to retrieve the information that remains in the autocorrelation of the measured voltage $V(t)$. Indeed, the qubit state dependent Kerr terms in $K_g$ and $K_e$ should allow for extra information to be retrieved in this way~\cite{Atalaya2018}. Finally, one could use an ancillary resonator dispersively coupled to the qubit to measure and demonstrate the entanglement between qubit and readout mode along their joint evolution as in Ref.~\cite{Vlastakis2015}.

\begin{acknowledgements} We thank Zaki Leghtas, Rapha\"el Lescanne, Mazyar Mirrahimi, Pierre Rouchon, Alain Sarlette, Hubert Souquet-Basiege, Matthias Droth, Marco Marciani, and Alexander Korotkov for fruitful interactions over the course of this project. The device was fabricated in the cleanrooms of Coll\`ege de France, ENS Paris, CEA Saclay, and  Observatoire de Paris. The Traveling Wave Amplifier was provided by the team of Will Oliver at Lincoln labs. The project was partly supported by Agence Nationale de la Recherche under project ANR-14-CE26-0018 and by the European Union’s Horizon 2020 research and innovation programme under grant agreement No 820505.
\end{acknowledgements}



\pagebreak
\widetext
\begin{center}
\textbf{\large Supplemental Materials: Sequential dispersive measurement of a superconducting qubit}
\end{center}
\setcounter{equation}{0}
\setcounter{figure}{0}
\setcounter{table}{0}
\setcounter{page}{1}
\makeatletter
\renewcommand{\theequation}{S\arabic{equation}}
\renewcommand{\thefigure}{S\arabic{figure}}
\renewcommand{\bibnumfmt}[1]{[S#1]}
\renewcommand{\citenumfont}[1]{S#1}

\section{Measurement setup}

\subsection{Microwave setup}
\begin{figure}[h!]
\includegraphics[scale=1.0]{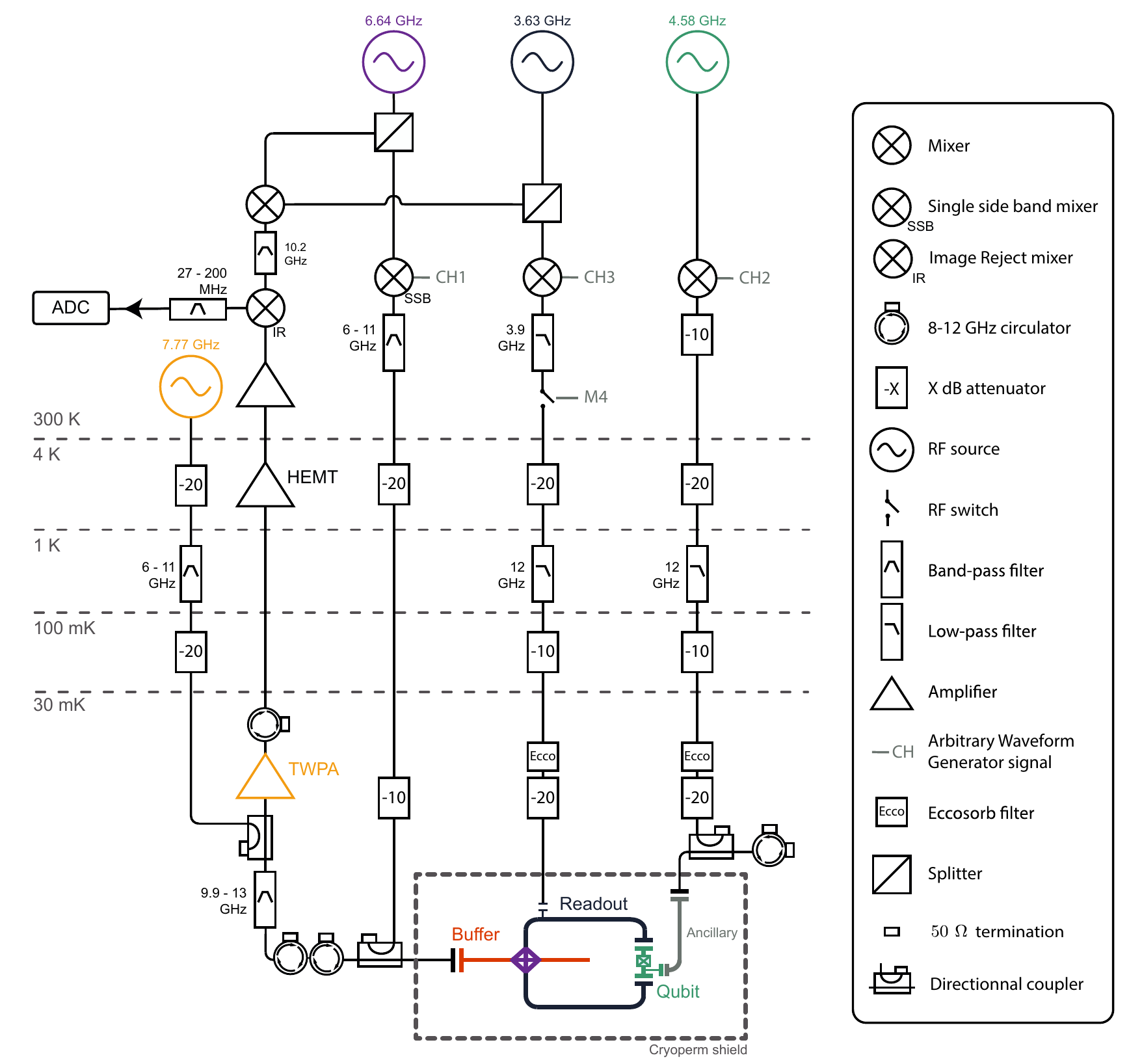}\caption{Wiring diagram. The RF source colors refer to the frequency of the matching element in the devices up to the modulation frequencies. The unused ancillary $6.3\ \textrm{GHz}$ resonator is coupled to a transmission line at $\kappa_\textrm{anc} = 2\times 2\pi \ \textrm{MHz}$. The $9.9-13 \ \textrm{GHz}$ band-pass filter is the combination of a high pass filter ($20 \ \textrm{cm}$ long WR62 waveguide) and a discrete low pass filter (Minicircuits\textregistered).}
\label{cablage}
\end{figure}

  The sample is cooled down to $T \approx 30~\mathrm{mK}$ in a BlueFors\textregistered dilution refrigerator. The scheme of the microwave input and output lines is provided  in Fig.~\ref{cablage}. The readout, pump and qubit pulses are generated by modulation of continuous microwave tones produced respectively by AnaPico\textregistered (APSIN12G), Agilent\textregistered (E8752D)and AnaPico\textregistered (APSIN12G) sources set at the frequencies  $ f_r - 100\ \mathrm{MHz}$, $f_b - f_r + 150 \ \mathrm{MHz}$ and $f_q+128 \ \mathrm{MHz}$. The pump tone is modulated through a single sideband mixer and the readout and qubit tones are modulated through regular mixers. The parasitic sidebands at the output of the mixers are far detuned from any frequency of interest and are neglected. The RF modulation pulses are generated by a 4 channels Tektronix\textregistered Abitrary Waveform Generator (AWG5014C) with a sample rate of $1 \ \mathrm{GS/s}$. Those pulses are of frequency $100\ \mathrm{MHz}$, $150\ \mathrm{MHz}$ and $128\ \mathrm{MHz}$. In order to ensure phase stability during the measurement, the local oscillator used for the demodulation of the output signal is not independent from the other ones. It is generated by mixing the output of the two sources that are close to the readout and pump frequencies, followed by the filtering out of the lower sideband. We thus retrieve a tone at frequency $(f_r - 100\ \mathrm{MHz}) +(f_b - f_r + 150 \ \mathrm{MHz}) = f_b + 50\ \mathrm{MHz} $ which is used to mix the signal down to $50\ \mathrm{MHz}$ before digitization.

The signal coming out of the buffer mode is filtered using a wave-guide with a cutoff frequency at $9.8\ \mathrm{GHz}$ in order to protect the following amplification stage to be affected by the strong reflected pump tone. The signal is first amplified by a Traveling Wave Parametric Amplifier (provided by W. Oliver group at Lincoln Labs). We tuned the pump frequency ($f_\mathrm{TWPA} = 7.773 \ \mathrm{GHz}$) and power in order to reach a gain of $19\ \mathrm{dB}$ at $10.220 \ \mathrm{GHz}$. The followup amplification is performed by a High Electron Mobility Transistor (HEMT) amplifier (from Caltech) at $4 \ \mathrm{K}$ and by two room temperature amplifiers.

\subsection{Sample description}

The sample was fabricated on a substrate of undoped Si (111) of dimension $8.67 \times 8.16 \times 0.28 \ \mathrm{mm}$. The ground planes and resonators are made of $150\ \mathrm{nm}$ of sputtered Nb after HF treatment. An optical lithography is performed using a laser writer before dry etching with SF6 the Nb layer. The Josephson junctions of the transmon and the JRM are fabricated using electronic lithography followed by an angle deposition of Al/AlOx/Al in a Plassys MEB550S evaporator. A good contact between Al and Nb is ensured by ion milling the Nb oxide in the Plassys evaporator prior to the evaporation over the overlap pads of $50 \times 50 \ \mu \textrm{m}^2$. The Nb sputtering was done in the Quantronics group at CEA Saclay, the wafer dicing at Observatoire de Paris, the  Al evaporation in a Plassys\textregistered ebeam evaporator and the laser writing at College de France, the electronic lithography at ENS Paris and the HF treatment at Paris Diderot. All measurements were performed at ENS de Lyon.

\begin{figure}[h!]
\includegraphics[width=\textwidth]{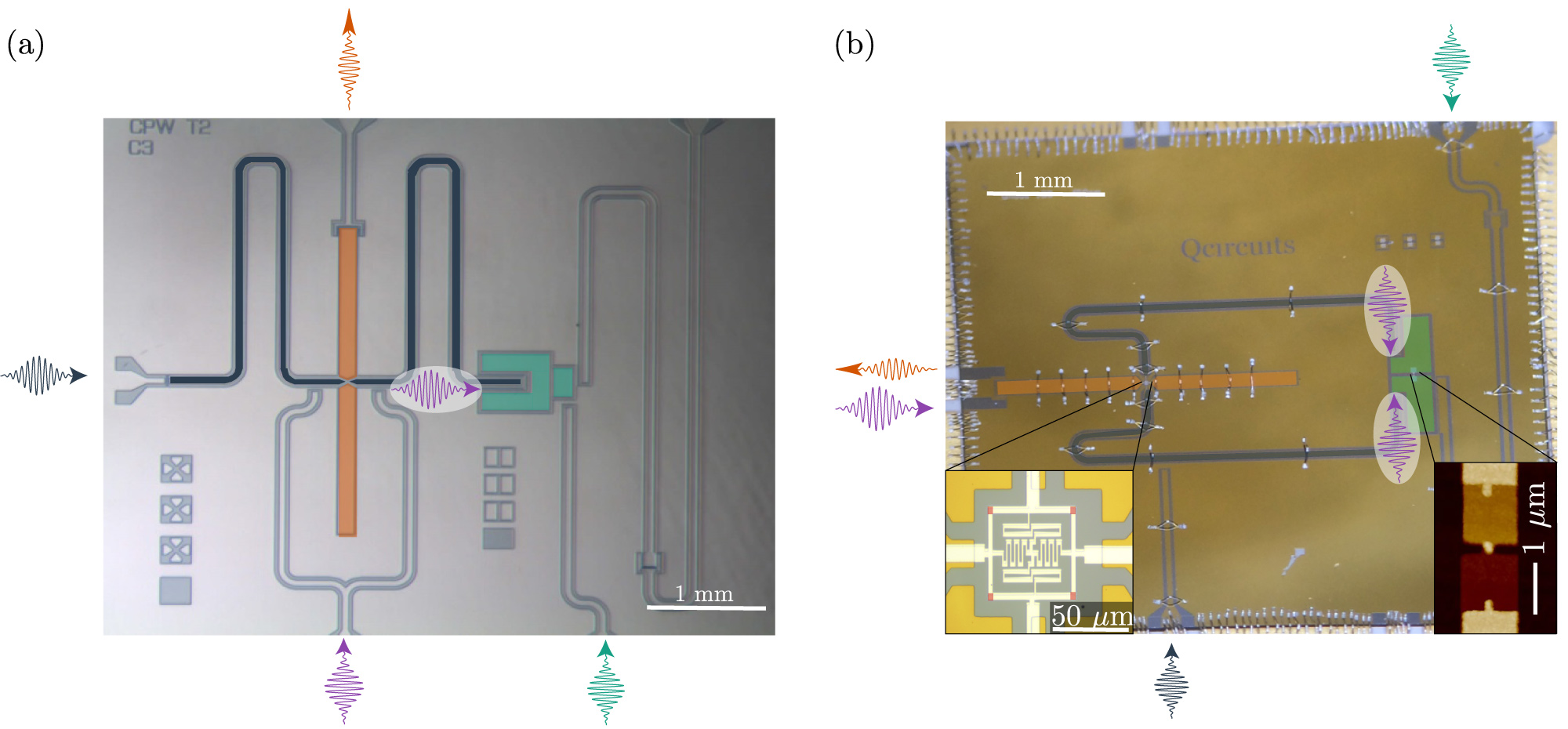}\caption{False color optical images comparing a previous design where the pump (purple) parasitically directly drives the transmon dipole (left) and the current design where this parasitic driving is avoided by symmetry (right). As in Fig.~1, the buffer is in orange, the readout resonator in dark gray and the transmon in green. Insets show the JRM (left) and single junction of the transmon qubit (right).}
\label{new_vs_old_design}
\end{figure}

The readout and the buffer are $\lambda / 2$ resonators in coplanar waveguide architecture. They are coupled in their middle by the JRM. The readout resonator has a gap width of $50 \ \mu \textrm{m}$ and a central conductor width of $84 \ \mu \textrm{m}$ whereas the buffer gap width is $10 \ \mu \textrm{m}$ and its central conductor width is $220 \ \mu \textrm{m}$. This design is intended to minimize the buffer characteristic impedance (here $25~\Omega$) in order to maximize the participation ratio of the JRM in that mode, without compromising too much on the reproducibility of the fabrication. The readout resonator is symmetrically coupled (see Fig.~1c) to the transmon qubit. This is a trick that enables the device to withstand large enough pump powers without affecting the transmon qubit. Indeed, the $\lambda/2$ resonator that acts as the readout mode provides an electric field that couples directly to the transmon dipole. However, the pump populates the common mode between the two resonators (Fig.~\ref{fig:modes}c) and which develops the same voltage on each end of the the $\lambda/2$ resonator and develop no voltage difference between the antennas of the transmon qubit. Hence, the pump induces negligible current across the transmon junction. This symmetric design thus reduces the coupling of the qubit to the pump drive by an estimated $20\ \textrm{dB}$ compared to an unpublished previous design (Fig.~\ref{new_vs_old_design}). This improvement was determined by measuring the minimal pump power that leads to ionization of the transmon similarly to what we have done in Ref.~\cite{Slescanne2018dynamics}.

\subsection{Device Hamiltonian}

\begin{figure}[h!]
\includegraphics[scale=1.0]{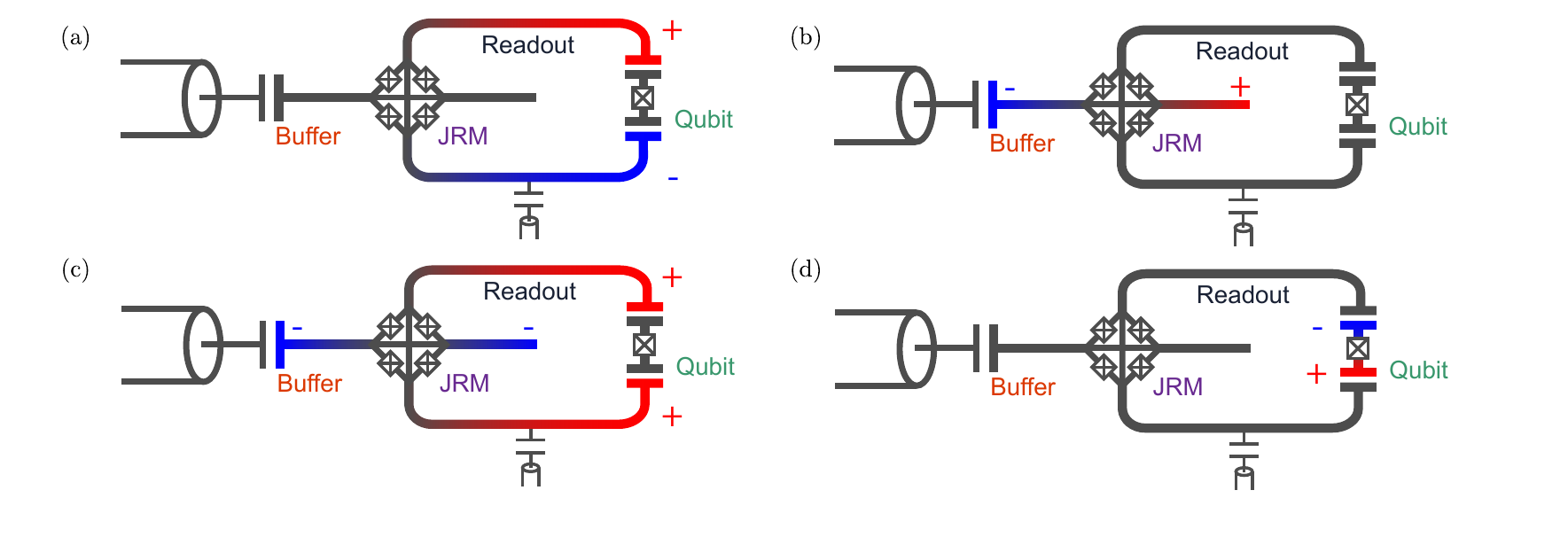}\caption{Scheme of the resonant modes that are used in the experiment. The color indicates the voltage as a function of position at maximum field amplitude in an oscillation. The readout mode (a) and the buffer mode (b) are $\lambda/2$ resonators. The common mode (c) extends over both CPW lines and does not couple to the transmon mode (d).}
\label{fig:modes}
\end{figure}

The full device Hamiltonian reads
\begin{equation}
    \hat{\mathcal{H}}=\hat{\mathcal{H}}_\mathrm{0}+\hat{\mathcal{H}}_\mathrm{JRM}+\hat{\mathcal{H}}_\mathrm{int},
\end{equation}
where the interaction free Hamiltonian is
\begin{equation}
   \hat{\mathcal{H}}_\mathrm{0} = \hbar \omega_{\rm b} \hat{b}^\dag \hat{b} + \hbar \omega_{\rm r} \hat{r}^\dag \hat{r} + \hbar\frac{\omega_{\rm q}}{2} \hat{\sigma}_z,
\end{equation}
the JRM Hamiltonian at the optimal flux point is~\cite{Sflurin:tel-01241123}
\begin{equation}
   \hat{\mathcal{H}}_\mathrm{JRM} = \hbar g_3 (\hat{r}+\hat{r}^\dag)(\hat{b}+ \hat{b}^\dagger)(\hat{c}+\hat{c}^\dagger),
\end{equation}
The mode $c$ is the common mode of the device (Fig.~\ref{fig:modes}c) and is driven by the pump. The dispersive interaction term between the qubit and the readout mode reads (see for instance Ref.~\cite{SAtalaya2018})
\begin{equation}
\begin{array}{cccl}
\hat{H}_{\textrm{int}} / \hbar = -\chi \hat{r}^\dagger \hat{r}|e\rangle \langle e|- K_g \hat{r}^{\dagger 2} \hat{r}^2 |g\rangle \langle g|- K_e \hat{r}^{\dagger 2} \hat{r}^2 |e\rangle \langle e|.
\end{array}
\end{equation}

When driven by the pump at $\omega_p=\omega_b-\omega_r$, the off-resonant common mode is occupied by a coherent state such that we can replace $\hat{c}$ by a scalar number $p$~\cite{Sflurin:tel-01241123}. In the rotating wave approximation, only two terms remain in $\hat{\mathcal{H}}_\mathrm{JRM}$, which then reads
\begin{equation}
   \hat{\mathcal{H}}_\mathrm{JRM} = \hat{H}_{\mathrm{bs}} = \hbar (g \hat{b}^\dagger \hat{r} + g^* \hat{b} \hat{r}^\dagger),
\end{equation}
where $g=pg_3$ is proportional to the pump amplitude $p_\mathrm{in}$.

\section{Calibration}

\subsection{Calibration of the quadratures and photon number of the readout mode}

The calibration of the readout mode quadratures in the Wigner functions and of the photon number in the readout mode were done by monitoring the free temporal evolution of the occupations of the first Fock states. To do so, we apply a readout mode displacement square pulse of $10\ \mathrm{ns}$ (or $20 \ \mathrm{ns}$ for Wigner tomography) with a $2 \ \mathrm{ns}$ Gaussian edge of a given amplitude to be calibrated in units of inner readout mode quadratures. We then wait for a time $t$ to let the readout mode decay before applying a selective $\pi$ pulse to the qubit at the frequency $\omega_q$ (resp. $\omega_q - \chi$). Hence we only excite the qubit if there is $0$ photon (resp. $1$ photon) in the readout mode. The selective $\pi$ pulse has a temporal envelope in $1/\cosh(\sqrt{\pi/2}t/\sigma)$ with a spread $\sigma = 200\ \textrm{ns}$. We assume a constant decay rate and an initial coherent state $|\alpha\rangle = \left|\sqrt{\Bar{n}}\right\rangle $. Therefore the probability to excite the qubit follows

\begin{equation}
\left\lbrace
\begin{array}{ccl}
P_{|0\rangle}(|e\rangle) &=& e^{-\Bar{n}e^{-\kappa_r t}}\\
P_{|1\rangle}(|e\rangle) &=& \Bar{n}e^{-\Bar{n}e^{-\kappa_r t}}e^{-\kappa_r t}
  \end{array}\right.
\end{equation}

\begin{figure}[h!]
\includegraphics[scale=1.0]{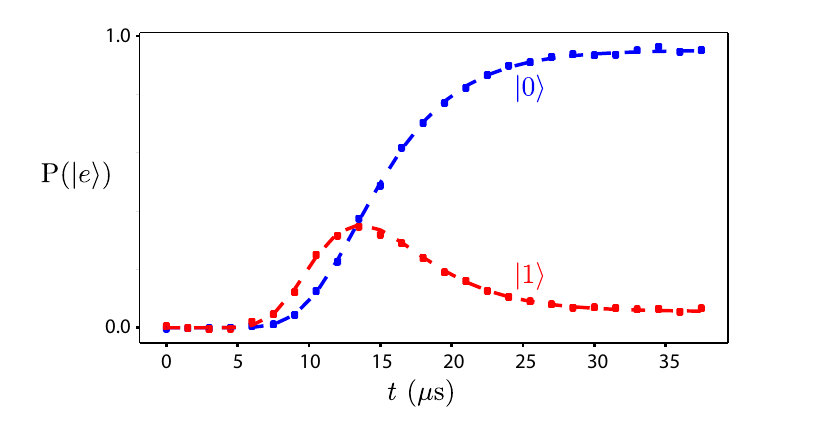}
\caption{Example of photon number calibration by monitoring the readout state during decay. Squares: Measured occupation probabilities of Fock states $|0\rangle$ and $|1\rangle$. Dashed lines: theory prediction with $\kappa_r = 1/T_\textrm{r} = 250 \times 2\pi \ \textrm{kHz}$ and initial photon number $\Bar{n} = 31.8$. Red corresponds to a selective $\pi$ pulse for 1 photon in the readout mode, blue to 0 photon.}
\label{photocomptage}
\end{figure}

In Fig.~\ref{photocomptage} are shown the measured occupations in the $|0\rangle$ and $|1\rangle$ readout states as a function of time. The data are  reproduced by the above equations provided the decay rate of the readout resonator is $\kappa_r = 1/T_\textrm{r} = 250 \times 2\pi \ \textrm{kHz}$ and the initial average number of photons $\Bar{n}$ is 31.8. Note that the theoretical curves (dashed lines) take into account a 0.95 scaling factor for $P_{|1\rangle}(|e\rangle)$ due to a slight imperfection in the selective $\pi$ pulse at $\omega_q - \chi$. We have repeated this procedure for each displacement amplitude and found the same $\kappa_r$ for all of them. This procedure provided a calibration between displacement amplitude and photon number.

\subsection{Characterization of the cavity pull and of the Kerr nonlinearities}

The measurement of $\chi$, $K_g$ and $K_e$ was done by monitoring the average phase acquired by the readout field as a function of time depending on the state of the qubit and the average photon number. 

\begin{figure}[h!]
\includegraphics[scale=1.0]{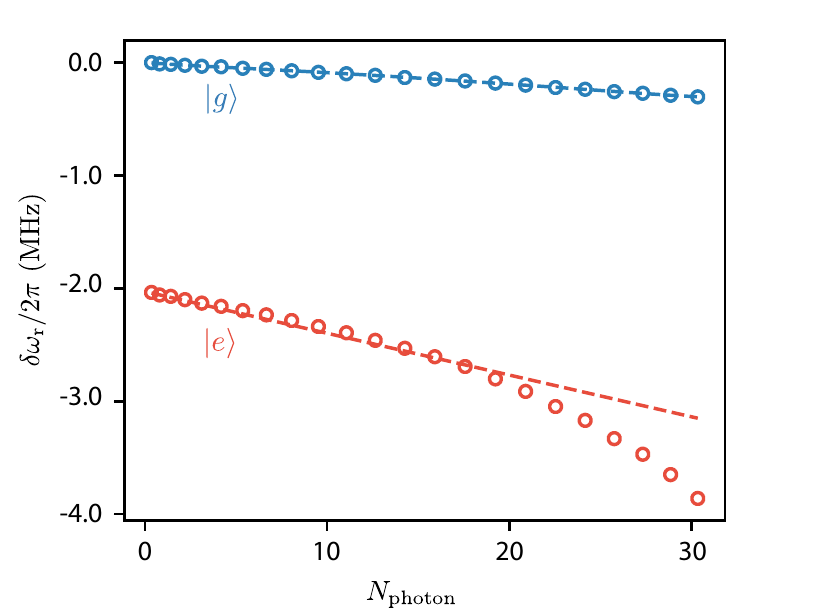}\caption{Average detuning of the readout resonator as the function of the average photon number and the state of the qubit. Data are cercles and first order fits are dashed lines. As the number of photons increases, the detuning deviates from the linear fit. Hence for the experiment displacement $\alpha = 5.8$ our simple model provides only a qualitative explanation. }
\label{detuningrate}
\end{figure}

Having prepared the qubit in either $|g\rangle$ or $|e\rangle$, we load the readout mode with a coherent state of amplitude $\alpha = \sqrt{\Bar{n}}$. We wait for a time $t_\mathrm{int}$. We then release the state of the readout mode into the transmission line and record the average phase $\phi(t_\mathrm{int})=\mathrm{arg}(\overline{\beta})$ of the released pulse. The detuning $\delta \omega_r$ between the resonant frequency of the readout resonator and a reference resonant frequency (when the readout resonator is in the vacuum state and for a qubit in $|g\rangle$) can be determined as $\delta \omega_r = \dfrac{d\phi}{dt_\mathrm{int}}$. In Fig.~\ref{detuningrate} are shown the measured detuning as a function of photon number in the readout resonator and of qubit state.

Our model for the qubit-readout resonator system in the rotating frame of the readout mode when the qubit is in the ground state, reads
\begin{equation}
\begin{array}{cccl}
\hat{H}_{\textrm{int}} / \hbar = -\chi \hat{r}^\dagger \hat{r}|e\rangle \langle e|- K_g \hat{r}^{\dagger 2} \hat{r}^2 |g\rangle \langle g|- K_e \hat{r}^{\dagger 2} \hat{r}^2 |e\rangle \langle e|.
\end{array}
\label{hamiltonian}
\end{equation}
The cavity pull $\chi=2.05\times 2\pi~\mathrm{MHz}$ can thus be read out as the readout frequency difference between the cases where the qubit is in $|g\rangle$ or $|e\rangle$, when the cavity is in the vacuum. The Kerr terms induce a linear dependence of the readout frequency as a function of average photon number. The Kerr rates $K_g$ and $K_e$ can thus be obtained from the slopes of the curves in Fig.~\ref{detuningrate}. At large photon number, a higher order polynomial fit is required to take into account higher order non-rotating terms of the Hamiltonian arising from the expansion of the Josephson cosine potential. As seen in the spiraling shape of the measured Wigner functions (Fig.~2), the readout mode displays much less anharmonicity when the qubit is in the ground state than in the excited state. 

\begin{table}
\centering
\begin{tabular}{c|c}
\hline
$\chi/2\pi$ & $2.05~\mathrm{MHz}$\\
\hline
$K_g/2\pi$ & $8.4~\mathrm{kHz}$ \\
\hline
$K_e/2\pi$ & $37~\mathrm{kHz}$ \\
\hline
\end{tabular}
\label{SecondOrderFit}
\end{table}

\subsection{Conversion rate and calibration of the coupling rate}

\begin{figure}[h!]
\includegraphics[scale=0.9]{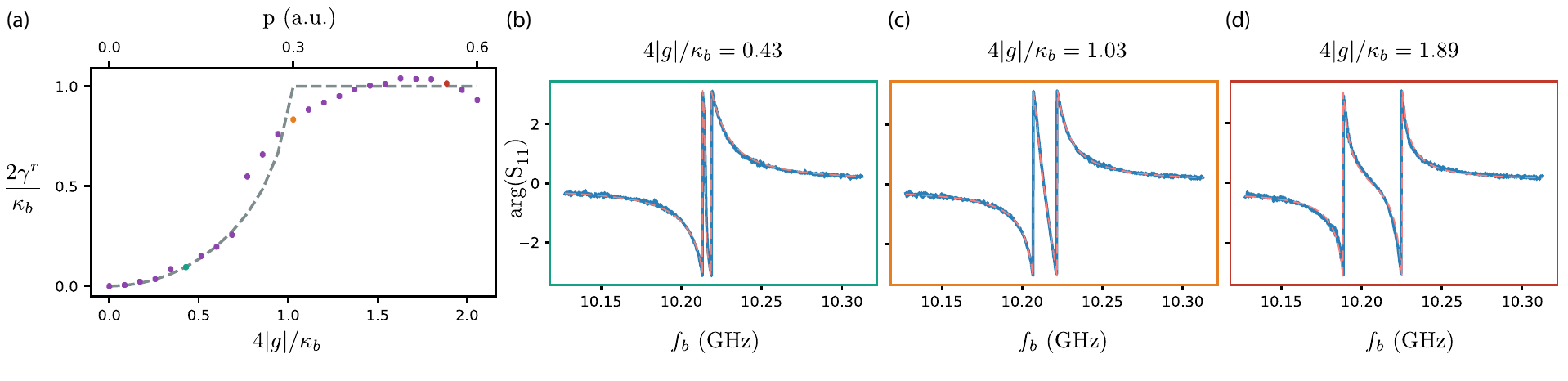}
\caption{(a) Direct coupling measurement (purple square) to the readout mode as a function of the pump amplitude. We obtain the calibration for the coupling g from the fit (dashed gray line). (b) to (d) Blue solid line : phase of the measured reflection coefficient on the buffer port for various pump amplitudes. Red dashed line : corresponding curve obtained by tuning $\gamma^r$, which allows us to plot one dot in (a). The color of each frame corresponds to the color of its corresponding dot in figure (a).}
\label{CWcoupling}
\end{figure}

By measuring the reflection coefficient on the buffer port as a function of frequency, one can determine the steady conversion rate $\gamma^r$ between the readout mode and the transmission line directly (see Eq.~(216) in Ref.~\cite{Sflurin:tel-01241123}). We have realized this measurement for various pump powers and obtained the measured dependence of $\gamma^r$ on pump amplitude (Fig.~\ref{CWcoupling}).

This curve can then be used to calibrate how the coupling rate $g$ that enters in the beam-splitter Hamiltonian depends on pump power. The 3-wave mixing Hamiltonian of the JRM indeed implies a linear dependence between the coupling rate and the pump amplitude $g(t) \propto p(t)$. In our limit $\kappa_b \gg \kappa_r$, the conversion rate indeed reads~\cite{Sflurin:tel-01241123}
 \begin{equation}
\gamma^r = \frac{\kappa_b}{2} \textrm{Re}\left [ 1- \sqrt{1-16\frac{|g|^2}{\kappa_b^2}} \right ]
\end{equation}
We observe that this model only faithfully reproduces the measurements for small pump powers such that $4|g|/\kappa_\textrm{b}<0.7$.

\begin{figure}[h!]
\includegraphics[scale=0.85]{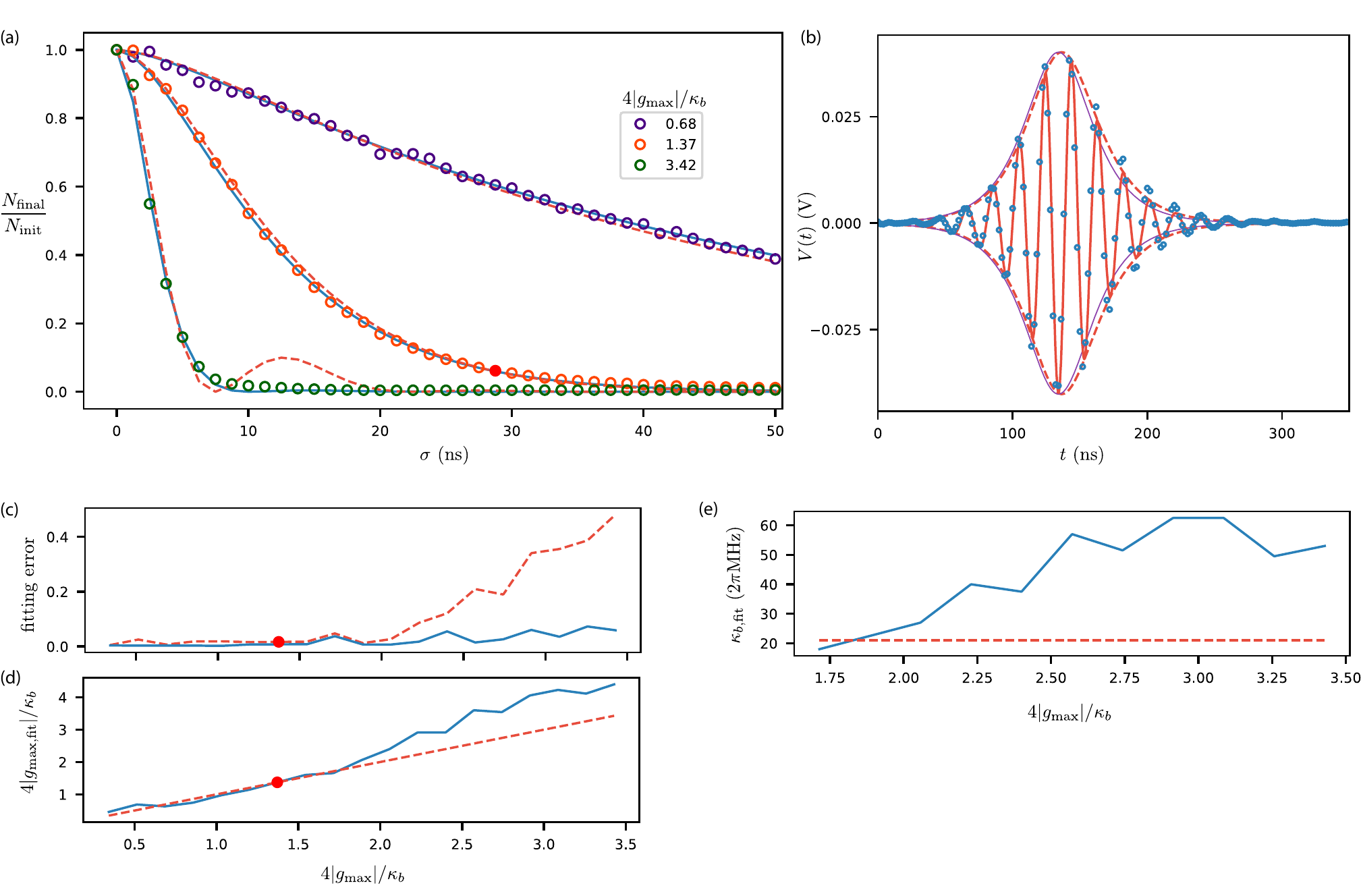}\caption{
(a) Fraction of photons remaining in the readout mode as a function of the spread $\sigma$ of the pump pulse for three pump powers. As in the other panels, the red dot corresponds to the release parameters of the main text. The initial number of photons is $N_\textrm{init} = 28$. Circles: measured photon number fraction through cavity state decay (see Fig.~\ref{photocomptage}). Red dashed lines: solutions of (\ref{eqmem}) using the calibrated values of $\kappa_b$ and $g_\mathrm{max}$ without fitting parameters. Discrepancies appear for fast flushes between the model and the experiment. We note that an empirical solution can be obtained by effectively increasing $\kappa_b$  and $4|g_\mathrm{max}|/\kappa_b$ beyond their calibrated values (blue solid lines).
(b) Dots: average recorded signal at the output of the final mixer on the buffer output line when the qubit is in $|g\rangle$, corrected for a DC offset (same as in Fig.~3b). Dashed red line: solution of \ref{eqmem} for the buffer amplitude $|b|$. An overall scaling factor is applied as well as an unique offset for the origin of time. Solid red line: same curve with a $51\ \textrm{MHz}$ modulation ($50~\mathrm{MHz}+(\chi/2)/2\pi$ because the qubit is in $|g\rangle$) and a fitted initial phase. A pump pulse envelope with the same amplitude is displayed in thin purple line for comparison.
(c) Relative fitting error (distance between the curves in (a) and the measured fraction of photon number) as a function of the pump amplitude in units of the independently determined value of  $4|g_\mathrm{max}|/\kappa_b$. Dashed red line: theory without free parameter. Solid blue line: theory with effective fitted values of $\kappa_b$ and $g_\mathrm{max}$. The red dot corresponds to the parameters of the main text.
(d) Corresponding effective values of $4|g_\textrm{max}|/\kappa_b$ as a function of the calibrated ones. The dashed line has a slope $1$.
(e) Corresponding effective values of $\kappa_b$ as a function of the independently calibrated $4|g_\textrm{max}|/\kappa_b$  for pump powers where the theory does not match the experiment. The effective increase of $\kappa_b$ can be interpreted as conversion into extra parasitic modes.}
\label{FastFlush}
\end{figure}
\subsection{Dynamics of the readout release}

The flushing of the readout mode by the smooth pump pulse is characterized as follows. The readout mode is loaded with a coherent state with amplitude $\alpha = 5.3$. In the following the qubit is assumed to be in its ground state\footnote{We did not measure the flushing of the readout mode when the qubit is excited}. It is justified by the fact that $\chi\ll\kappa_\textrm{b}$, for which the release dynamics is similar for the qubit in the ground or in the excited state. A pump pulse with an envelope $p(t) \propto \left(\cosh{(\sqrt{\pi/2} t/ \sigma)}\right)^{-1}$ is applied for various pump amplitudes and spread (Fig.~\ref{FastFlush}). This choice of a smooth pump pulse allows us to limit the spectral broadening of the pump. The Fourier transform of that pulse has the same shape but with a variance $1/\sigma$. In practice, the pump pulse is windowed by a square shaped weight function of duration $8\sigma$. We then measure the remaining average photon number using the technique of cavity state decay as in Fig.~\ref{photocomptage}. 

The release dynamics is captured by the quantum Langevin equation for the field amplitudes $m$ and $b$ in the rotating frames of the buffer and the readout mode. 

\begin{equation}
\left\lbrace
\begin{array}{ccl}
\Dot{r} & = &- ig^\star b -\frac{\kappa_r}{2}r\\
\Dot{b} & = &-igr - \frac{\kappa_b}{2}b 
\end{array}\right.
\end{equation}

Hence,

\begin{equation}
\left\lbrace
\begin{array}{ccl}
\Ddot{r} + \left( \frac{\kappa_b}{2}+\frac{\kappa_r}{2} - \frac{\Dot{g}^\star}{g^\star} \right)\Dot{r} + \left(|g|^2+\frac{\kappa_r \kappa_b}{4}-\frac{\kappa_r}{2}\frac{\Dot{g}^*}{g^*} \right)r &=&0 \\
g^\star b - i \left( \Dot{r} + \frac{\kappa_r}{2}r \right) &=&0
\end{array}\right.
\label{eqmem}
\end{equation}
with $g(t) = g_\textrm{max}/\cosh{\left(\sqrt{\pi / 2} t/\sigma\right)}$ and $\frac{\Dot{g}^\star}{g^\star}(t) = - \frac{\sqrt{\pi / 2} \tanh{\left(\sqrt{\pi / 2} t/\sigma\right)}}{\sigma}$. 

In Fig.~\ref{FastFlush}a are shown the measured readout photon number normalized to the initial number as a function of $\sigma$ and for various pump powers. The numerical solution of the above equation reproduces well these measurements as long as $4|g_\mathrm{max}|/\kappa_b$ does not exceed 1.6 (Fig.~\ref{FastFlush}c). There, even the average time trace of the buffer output is well captured (Fig.~\ref{FastFlush}b).

For the largest pump powers, for which this value is exceeded, the readout mode is flushed more rapidly than theoretically expected. The observed behavior can be captured by allowing to tune the effective values of $\kappa_b$ and $g_\mathrm{max}$. The increase of the effective decay rate $\kappa_\textrm{b}$ could be due to the conversion of the readout mode into parasitic unmonitored modes (Fig.~\ref{FastFlush}e).

\section{Signal processing}

\subsection{Demodulation basis construction}

As the readout mode is released into the buffer line, its quantum state is mapped into a propagating microwave mode. How to determine the quadratures of the propagating mode into which the quadratures of the inner readout mode are linearly mapped? Here, the signal is carried by a mode at $\omega_b$ when the qubit is in $|g\rangle$ and $\omega_b-\chi$ when in $|e\rangle$. However, the pulse takes a finite duration corresponding to the temporal extent of the pump pulse. The release operation thus leads to a frequency spread of $\Delta\omega=1/\sigma$. In our experiment, we have $\Delta\omega=1/\sigma\approx 2.7\chi$. 

Before addressing the general case, let us first disregard this frequency uncertainty due to the qubit state and discuss how to find the quadratures of the propagating mode. One way to determine them would be to measure the average voltage trace at the output of the last mixer in the buffer line detection setup. We denote $\overline{V_{|\alpha\rangle}(t)}$ the average trace when the readout mode is in a coherent state $|\alpha\rangle$. An example of this curve can be found in Fig.~\ref{FastFlush}b. The signal coming out of an image reject mixer is oscillating at about $50~\mathrm{MHz}$ so that its amplitude and phase match the ones of the mode close to $\omega_b$ at the input of the image reject mixer. The purpose of this downconversion is to be able to digitize the signal with an acquisition board. We could then construct a demodulation function 
\begin{equation}
w_1(t)=\overline{V_{|\alpha=1\rangle}(t)}+i\overline{V_{|\alpha=i\rangle}(t)}.
\end{equation}
We define the complex amplitude $\beta_1=\int{V_\alpha(t) w_1(t) dt}$ as in Eq.~(2) of the main text. Here $V_\alpha(t)$ is a single voltage trace measured when the readout is in $|\alpha\rangle$. Ideally, we want to have $\overline{\beta_1} = \lambda \alpha$. Scaling $w_1$ using the inverse proportionality factor $\lambda$ would then lead to $\overline{\beta_1}=\alpha$ as required.

In practice, dc measurement offsets and drifts occur and we can simply eliminate them by defining the weight function as
\begin{equation}
w_\mathrm{2}(t)=\frac{\overline{V_{|\alpha=1\rangle}(t)}-\overline{V_{{|\alpha=-1\rangle}}(t)}}{2\lambda}+i\frac{\overline{V_{{|\alpha=i\rangle}}(t)}-\overline{V_{|\alpha=-i\rangle}(t)}}{2\lambda}.
\end{equation}

It is also possible to avoid two measurements by calculating the imaginary part of $w(t)$ from its real part alone in the case of slowly varying temporal envelope compared to the modulation frequency. Indeed, the imaginary part corresponds to phase shifting the carrier by $\pi/2$ while preserving the signal envelope.
\begin{equation}
\mathrm{Im}[w(t)]=\mathrm{rFT}^{-1}\left[i\times \mathrm{rFT}\left(\mathrm{Re}\left[w(t)\right]\right)\right],
\end{equation}
where $\mathrm{rFT}$ and $\mathrm{rFT}^{-1}$ respectively are the real Fourier transform and the inverse real Fourier transform. We can then define the weight function from its real part alone as
\begin{equation}
    \mathrm{Re}\left[w_3(t)\right]=\frac{\overline{V_{|\alpha=1\rangle}(t)}-\overline{V_{{|\alpha=-1\rangle}}(t)}}{2\lambda}
\end{equation}

Now, in the general case, the presence of the qubit leads to two possible frequencies for the propagating mode. We thus need to use a slightly different weight function. The weight functions corresponding to the qubit in $|g\rangle$ or $|e\rangle$ can simply be averaged here. This choice would then lead to 
\begin{equation}
    \mathrm{Re}\left[w_4(t)\right]=\frac{\overline{V_{|\alpha=1\rangle\otimes|g\rangle}(t)}+\overline{V_{|\alpha=1\rangle\otimes|e\rangle}(t)}-\overline{V_{{|\alpha=-1\rangle\otimes|g\rangle}}(t)}-\overline{V_{|\alpha=-1\rangle\otimes|e\rangle}(t)}}{4\lambda}
\end{equation}
Since the dc offsets and drifts are assumed to be independent of the qubit state, two of these traces are redundant.
\begin{equation}
    \mathrm{Re}\left[w_5(t)\right]=\frac{\overline{V_{|\alpha=1\rangle\otimes|g\rangle}(t)}-\overline{V_{|\alpha=-1\rangle\otimes|e\rangle}(t)}}{2\lambda}
\end{equation}

In the end, we chose to use the measurement of the voltage (Fig.~3b) when $\alpha=5.8$ initially. In this case, waiting for an interaction time $t_\mathrm{int}=100~\mathrm{ns}$ leads to $\alpha\approx -5.8$ in the excited state and $\alpha\approx 5.8$ in the ground state. We thus define the final weight function as in the main text as
\begin{equation}
    \mathrm{Re}\left[w(t)\right]=\frac{\overline{V_{|\alpha=5.8\rangle\otimes|g\rangle}(t)}-\overline{V_{|\alpha=-5.8\rangle\otimes|e\rangle}(t)}}{2\lambda}~\textrm{ and }\mathrm{Im}[w(t)]=\mathrm{IrFT}\left[i\times \mathrm{rFT}\left(\mathrm{Re}\left[w(t)\right]\right)\right].
\end{equation}
One can show that the exact value of the field amplitude after the waiting time is unimportant as it simply modifies the complex $\lambda$ scaling factor.

In order to determine $\lambda$, we thus match the scaling of the weight function $w(t)$ so that the measured average complex amplitude $\overline{\beta}$ matches the complex amplitude $\langle \hat{r}\rangle$ in the readout mode when the interaction time is set to $t_\mathrm{int}=0$ and when the qubit is in $|g\rangle$.

\subsection{Total efficiency}

The total efficiency of the release and detection of the propagating mode can be inferred from the variance of the histogram of measured $\beta$'s for $10^5$ runs of the experiment, when the readout mode is in a coherent state. We find a total efficiency $\eta = 11\ \%$. This efficiency can be understood as the product of the efficiency of each step of the signal measurement and analysis. The incomplete release of the readout state induces an efficiency of $\eta_\textrm{release}=91\%$ (see red dot in Fig~\ref{FastFlush}a). The total efficiency is standard~\cite{SBultink2017}, and can be understood by the attenuation of the several commercial microwave components between the device and the TWPA, including a 20cm long waveguide acting as a filter (see Fig.~\ref{cablage}). The TWPA provided $19\ \textrm{dB}$ of gain and its stop-band frequency is $2.4\ \textrm{GHz}$ below the buffer signal frequency.

\section{Choice of parameters}

\subsection{Optimal release amplitude and temporal envelope}

The goal of this experiment is to demonstrate a sequential single shot read-out with the highest achievable fidelity. We chose the optimal parameters by  minimizing the overlap between the distribution $P_{|e\rangle}$ and $P_{|g\rangle}$. The pump temporal shape was chosen to minimize this overlap. There is a  trade-off between the speed of the flush and the measurement fidelity. Indeed, a fast flush reduces the efficiency as it dissipates a fraction of the readout energy in unmonitored modes (see Fig.~\ref{FastFlush}). On the other hand, a slower flush is not able to release the readout mode when it is at the maximum phase difference. Instead, it continuously releases the content of the readout mode as it interacts with the qubit.


In our experiment, this trade-off yields an optimal pump spread $\sigma = 28 \ \textrm{ns}$ and maximum coupling $|g_\textrm{max}| = 7.2\times 2\pi \ \textrm{MHz} > \frac{\kappa_\textrm{b}}{4}$ corresponding to a slight over-critical coupling of the readout resonator and the buffer. 

\subsection{Operating flux bias of the JRM}

\begin{figure}[h!]
\includegraphics[scale=1.0]{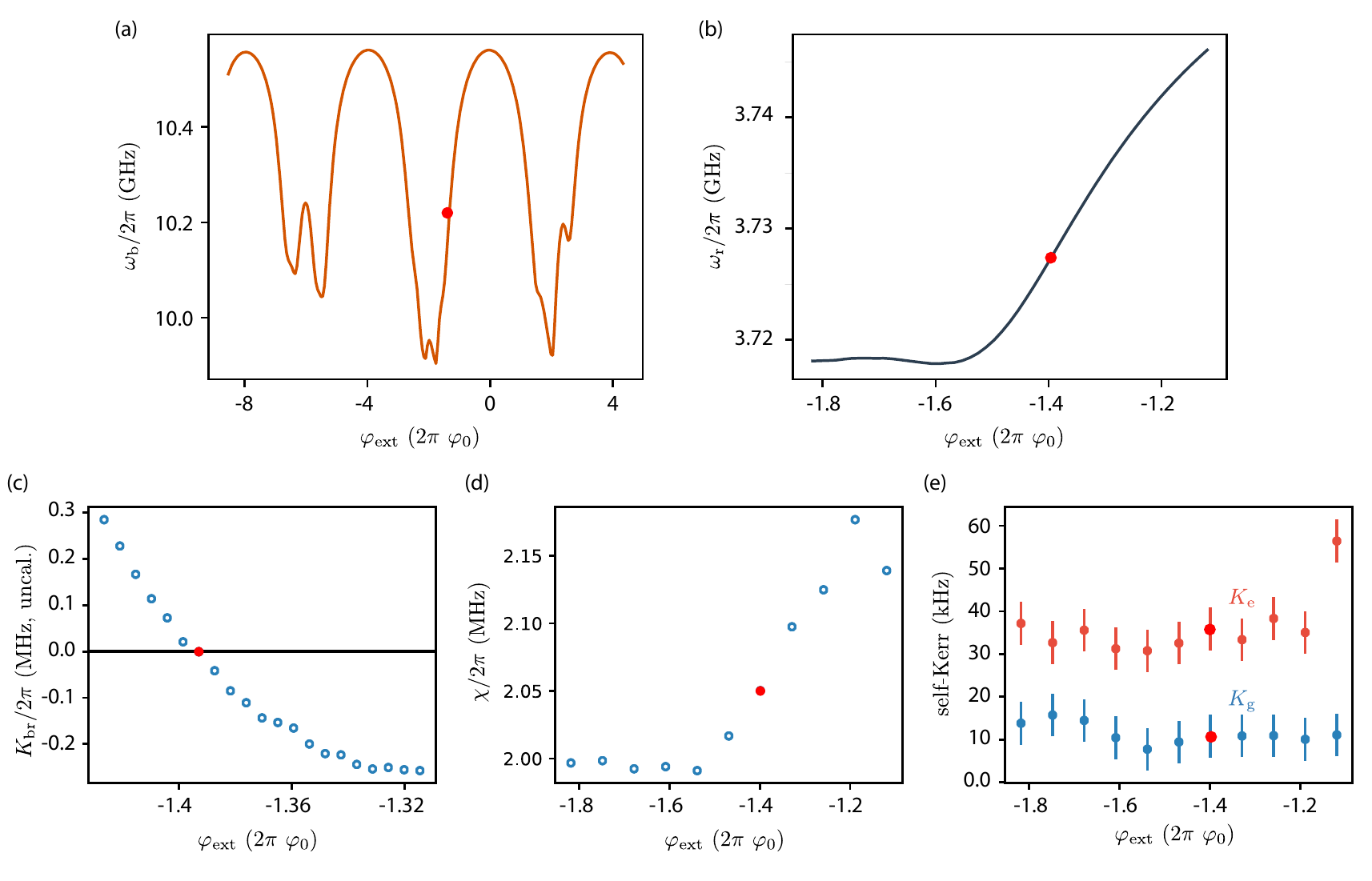}
\caption{Characterisation of the device as the function of external flux. The red dots correspond to the chosen flux point for the experiment in all panels. (a) Measured frequency of the buffer mode displaying a $4 \phi_0$ periodicity. The additional non periodic patterns are likely due to an asymmetry between the areas of the four subloops of the ring. (b) Frequency of the readout mode zoomed around the chosen operation point. (c) Frequency shift of the readout mode $K_{br}n_b$ induced by $n_b$ photons in the buffer. Contrary to other plots, this value is uncalibrated as $n_b$ is unknown (estimated $n_\textrm{b}\approx 10 ~$photons). Canceling out the cross-Kerr rate $K_{br}$ was our criteria for setting the operating flux bias. (d) $\chi$ as the function of the external flux. (e) Kerr rates for the qubit in $|g\rangle$ in blue and for the qubit in $|e\rangle$ in red as a function of the external flux.}
\label{Flux}
\end{figure}

The behaviour of the shunted JRM as a function of the external flux was studied in depth in previous works~\cite{Sflurin:tel-01241123}. By measuring the reflection coefficient on the buffer port, we determined the buffer frequency as a function of the external magnetic flux (Fig.~\ref{Flux}a). 

We then characterize the cross-Kerr effect induced by the buffer occupation on the readout resonator. To do so, we displace the readout mode with $\alpha_0 \approx 5$. Then we apply a continuous tone at the buffer frequency on the buffer port for a time $t$ corresponding to an estimated $10$ photons in the buffer at the end of the pulse, which is large compare to $1/\kappa_\textrm{b}$. Finally we release the readout mode state into the buffer and measure its average phase $\mathrm{arg}(\overline{\beta})$. The acquired phase as a function of  time $t$ provides us the readout mode frequency (as in Fig.~\ref{detuningrate}). We subtract to this measured frequency the one measured without driving the buffer and obtain a measurement of the modification of the readout mode frequency by the buffer occupation without any calibration of the buffer photon number at this stage. This cross-Kerr effect cancels out for a given value of the external flux as expected for the JRM. We chose this flux point for the experiment as it ensures that the readout mode is decoupled from the transmission line during the interaction time and as it also cancels out the cross-Kerr effect induced by the pump tone during the release. We use the same method as in Fig.~\ref{detuningrate} to measure the self-Kerr rate and $\chi$ as a function of the external flux. The self-Kerr rate is dominated by the contribution of the single junction transmon qubit and hence show little to no flux dependency. Increasing the JRM participation ratio should increase this Kerr tunability by the flux and enables a canceling of the average Kerr in the readout mode.

\newpage
\section{Wigner Tomography}

\subsection{Measurement}

The measurement protocol for the Wigner tomography matches that of previous work~\cite{SLutterbach1997,SBertet2002,SVlastakis2013,SBretheau2015}. To measure $W(\alpha)$ we start by applying a displacement $D(\alpha)$ on the readout mode with a $20 \ \textrm{ns}$ pulse at its frequency. Then we perform two $\pi/2$ unconditional pulses on the qubit\footnote{In practice, we interleave the pulse sequences to rotate the second pulse by $\pi/2$ or $-\pi/2$ and subtract their average outcome to eliminate parasitic drifts.} separated by a time $\pi/\chi = 236 \ \textrm{ns}$. We then flush the readout mode (same pulse width $\sigma=28~\mathrm{ns}$) and use our full readout protocol to measure the state of the qubit. The release step avoids the usual cross-Kerr effect between the readout and the readout mode that may distort the estimated Wigner function at large photon number and is mandatory for us as we use the readout resonator to measure the qubits state.

For the Wigner tomography starting with an excited qubit, we simply take the opposite of the result to account for the qubit being initially excited.

Each Wigner tomography is decomposed in 16 square panels spanning the acquired phase space that are each $20\times20$ pixels in size. Each pixel is averaged $5000 $ times. The obtained images are numerically rotated by $24^\circ$ for $|g\rangle$ (corresponding to setting $\mathrm{arg}(\lambda)=0$) and $97^\circ$ for $|e\rangle$. The latter choice accounts for a systematic angular offset between the initial displacement of $10 \ \textrm{ns}$ for the qubit readout experiment and the longer Wigner displacement of $20 \ \textrm{ns}$, which is hence more sensitive to the frequency shift $\chi$ of the readout mode.

\begin{figure}[!h]
\includegraphics[scale=0.9]{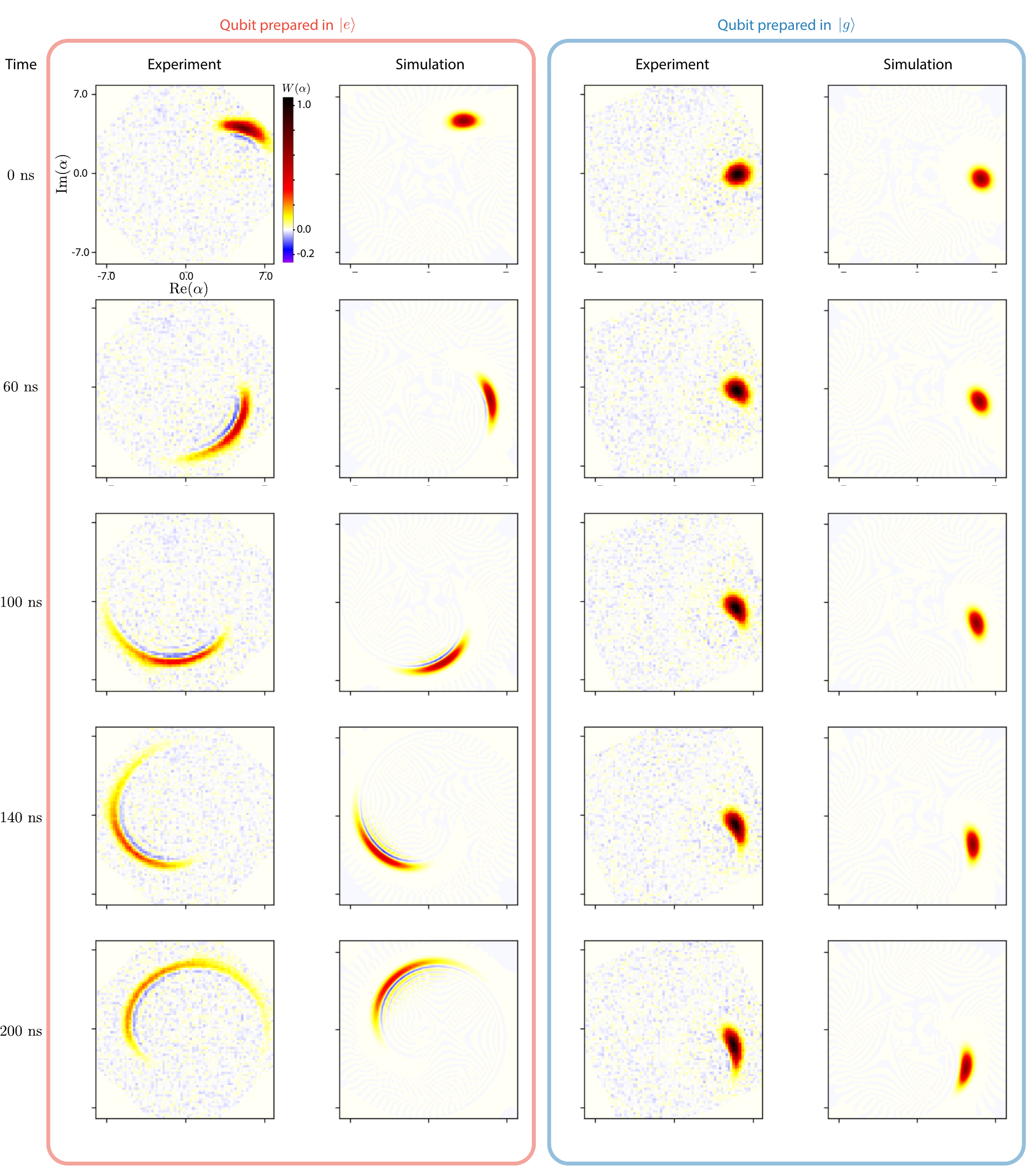}
\caption{Measured Wigner functions for various interaction times when $\alpha_0=5.8$ and their simulated counterparts displaying a qualitative agreement. The numerical simulations are done without any free parameters in a Hilbert space that is truncated to the first 150 energy levels (see text).}
\label{Wigners}
\end{figure}

\subsection{Numerical simulation of the Wigner function of the readout mode in time}
In order to validate our model for the evolution of the readout mode, we performed simulation of its dynamics using the Python package QuTip~\cite{SJohansson2013} with the Hamiltonian (\ref{hamiltonian}) and the independently measured values for its parameters (see above sections). The fit includes an extra $20\ \textrm{ns}$ delay to account for the duration of the displacement pulse. The results are provided on Fig.~\ref{Wigners} and display a qualitative agreement with the measurement. It confirms that the dynamics of the readout mode is mainly governed by the $\chi$ and Kerr terms of the Hamiltonian.

\section{Readout characterization}

\subsection{Measurement fidelity}

\begin{figure}[h!]
\includegraphics[scale=1.0]{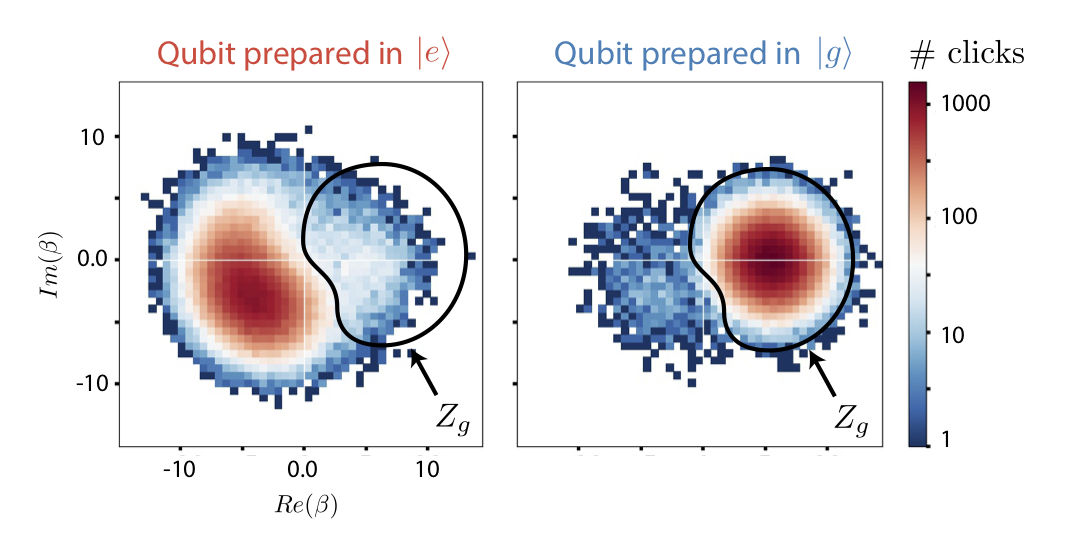}\caption{Recorded distribution of $\beta$ in logarithmic scale for the number of detection in each bin, when the qubit is prepared in $|e\rangle$ (left panel) or in $|g\rangle$ (right panel). An approximate contour of $Z_g$ is represented in black. $\mathcal{E}_g$ corresponds to clicks outside of $Z_g$ on the right panel and $\mathcal{E}_e$ to clicks inside of $Z_g$ on the left panel. }
\label{LogScaleHist}
\end{figure}

To extract a binary answer to whether the qubit is excited or not for a given realization of the experiment, we define two disjoint ensembles of possible outcomes $\beta$. The complex amplitudes corresponding to the answer '$g$' belong to $Z_g = \{ \beta \ : \ P_{|g\rangle}(\beta)>P_{|e\rangle}(\beta) \}$, which is defined from the measured probability densities. Obtaining $\beta \notin Z_g$ correspond to the result 'e' (see Fig.~\ref{LogScaleHist}). Given the distributions, the error probability to find an amplitude $\beta$ in $Z_g$ when the qubit was prepared in $|e\rangle$ is $\mathcal{E}_e = 3.4\ \%$ while the reverse error is $\mathcal{E}_g = 1.6\  \%$. They can be in part explained by the imperfect preparation of the states $|e\rangle$ and $|g\rangle$ because of gate infidelity ($0.5\ \%$) and qubit thermal population ($0.8\ \%$). Those are described below. The qubit decay during the interaction time accounts for $1.7 \ \%$ of the error in $\mathcal{E}_e$.

\subsection{Pulse fidelity estimation}

\begin{figure}[h!]
\includegraphics[scale=1.0]{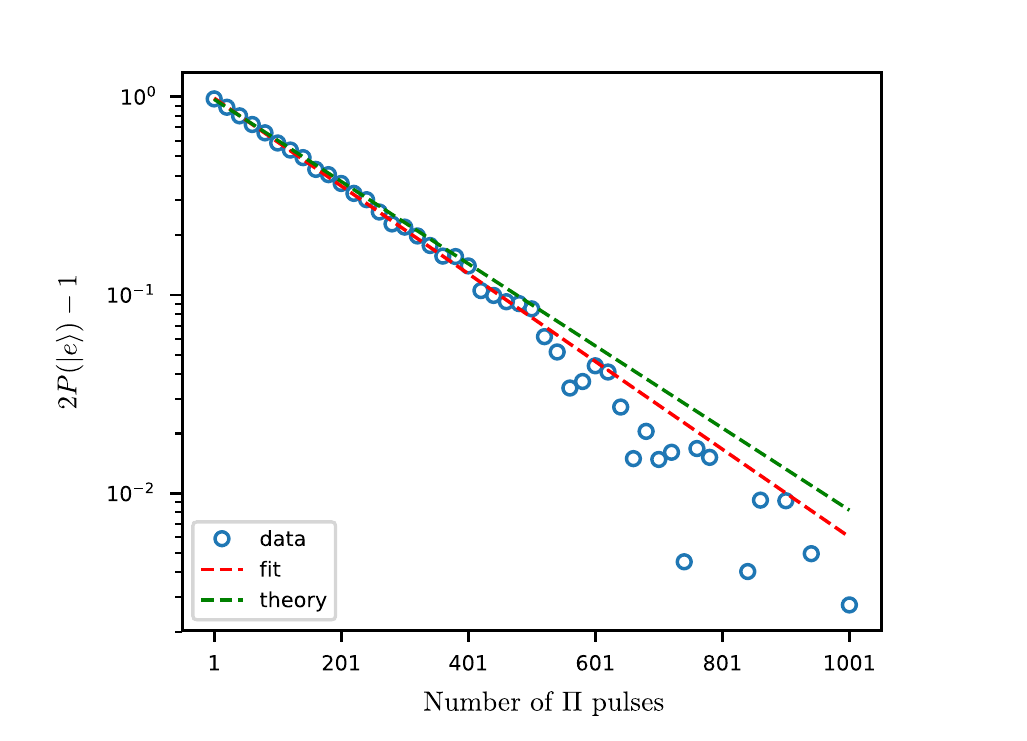}
\caption{ Measured population in the excited state of the qubit as a function of the odd number $N$ of $\pi$ pulses in series. Circles: measured probabilities. Red dashed line: exponential fit indicating a decrease by a factor $F=0.995$ between one pulse and the next. Green dashed line: predicted exponential decay of Rabi oscillations taking the same duration. The decay is $\propto \mathrm{exp}(-N \Delta t_p (T_1^{-1}+T_2^{-1})/2$ ~\cite{SFicheux2018}.}
\label{Fidelity}
\end{figure}

In order to determine the pulse fidelity, we perform an odd number $N$ of $\pi$ pulses in series and measure the probability to find the qubit in the excited state afterwards.  The probability decays exponentially with the number of pulses. From the decay, we determine a $99.5\ \%$ fidelity whose departure from $100~\%$ mostly comes from the decay and decoherence of the qubit during the pulse (Fig.~\ref{Fidelity}). Each $\pi$ pulse is $30 \ \textrm{ns}$ long and has a sech shape with a spread of $7.5 \ \textrm{ns}$ and the $\pi$ pulses are separated by a $5\ \textrm{ns}$ resting time amounting to $\Delta t_p=35~\mathrm{ns}$ of period between two pulses. 

\subsection{Qubit thermal population}

\begin{figure}[h!]
\includegraphics[scale=1.0]{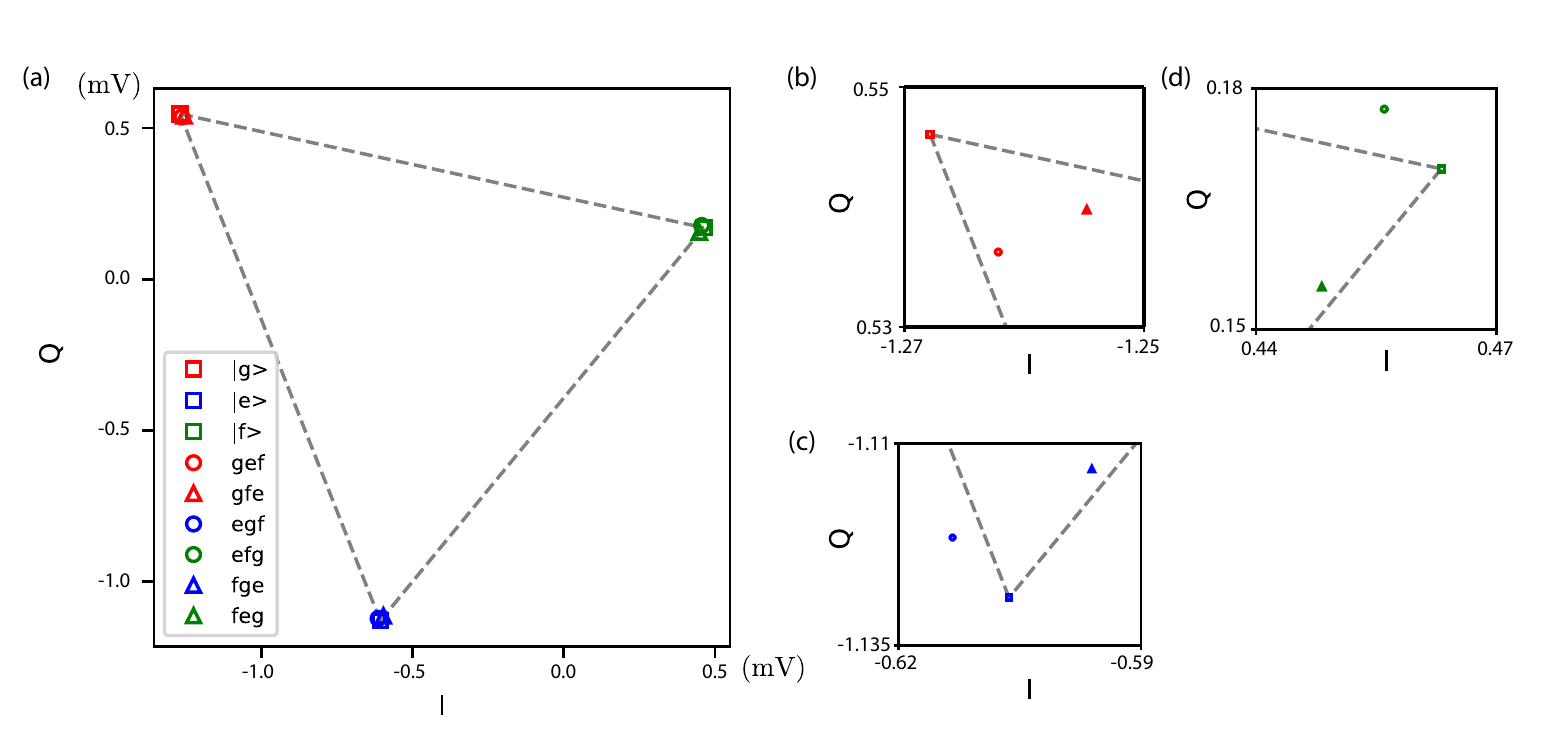}
\caption{Measurement of the transmon temperature. (a) Average response in the IQ plane of the readout mode depending on the population permutations. Squares correspond to the fitted readout response if the transmon were entirely in one of its first three lowest energy states. Other markers correspond to the average measured quadratures after the following pulse sequences (time ordered from left to right) : 'gef' - no pulse; 'gfe' - $\pi_{ef}$; 'egf' - $\pi_{ge}$; 'efg' - $\pi_{ge} \pi_{ef}$; 'fge' - $\pi_{ef} \pi_{ge}$; 'feg' - $\pi_{ge}\pi_{ef}\pi_{ge}$. Solving the associated linear problem yields a temperature of $44\ \textrm{mK}$ corresponding to $0.8\ \%$ of the qubit's population not in the ground state. Dashed lines are visual guides connecting the squares. (b),(c) and (d) are zoomed around the apexes. Data points should be inside of the triangle as they are a weighted sum of the apex, with weights given by the populations of each of the three states. The uncertainty on data point is about $0.01~\textrm{mV}$.}
\label{Thpop}
\end{figure}

The qubit thermal excitation was measured by using the 3rd level of the transmon qubit as a reference. Indeed, by exploring all configurations of population permutation between those 3 levels, one can retrieve the qubit thermal population distribution and its temperature (Fig.~\ref{Thpop})~\cite{SFicheux2018}. The system of linear equations corresponding to the readout response as a function of the population in each state of the transmon is solved through a least square minimization to account for the measurement uncertainty.  We find $T_{qubit} = 44\ \textrm{mK}$ corresponding to $0.8\ \%$ of the qubit population out of the ground state.

\subsection{Quantum Non Demolition nature}

\begin{figure}[h!]
\includegraphics[scale=1.0]{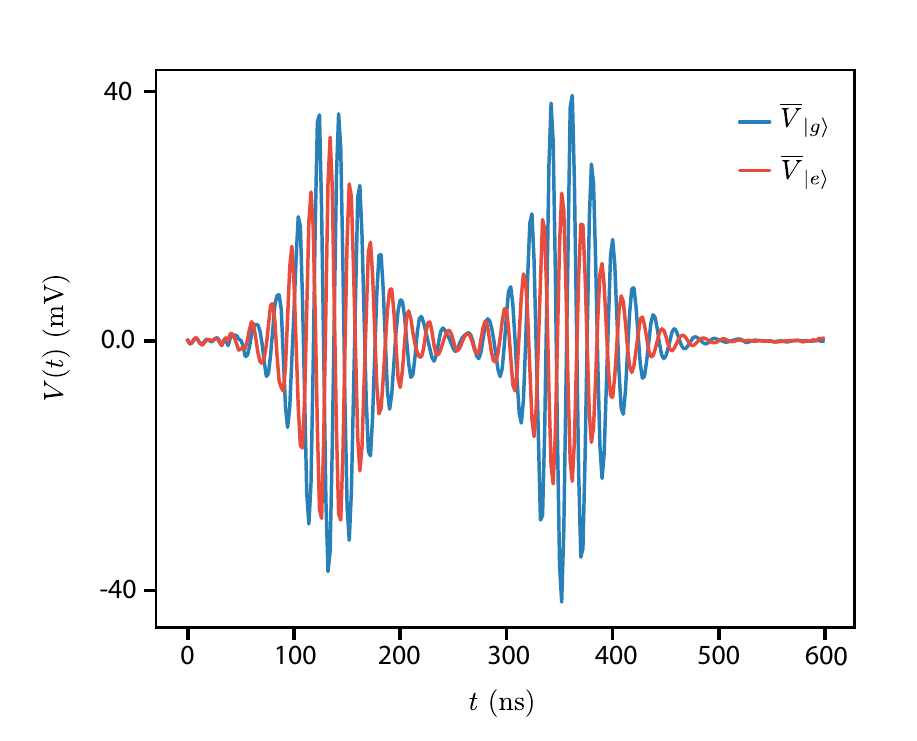}
\caption{Average measured signals for two consecutive readouts for a qubit prepared in $|g\rangle$ (blue) or in $|e\rangle$ (red). The second readout starts $220\ \textrm{ns}$ after the first one.}
\label{QDNness}
\end{figure}

The Quantum Non Demolition characteristic of the measurement (its so-called QNDness) was determined by doing two readouts, one right after the other (Fig.~\ref{QDNness}). By heralding on the outcome of the first readout for the realizations that give the expected outcome given the preparation, we measure a probability for the second readout to give the same outcome of $95\ \%$ . This slight departure from perfect QNDness is expected as the coupling relies on the dispersive approximation while the field amplitude in the cavity is $\alpha = 5.8$ and thus a photon number of 34, which is above the critical number of photons $n_\mathrm{crit}=(\Delta/2g)^2\approx\alpha/2\chi\approx 24$~\cite{SBlais2007,SKoch2007}.

\bibliographystyle{landry}
\bibliography{references}

\end{document}